\documentclass[11pt,a4paper]{article}
\usepackage{amssymb}
\usepackage{times,cite}
\newcommand{\be}{\begin{eqnarray}}
\newcommand{\ee}{\end{eqnarray}}
\newcommand{\bez}{\begin{eqnarray*}}
\newcommand{\eez}{\end{eqnarray*}}
\newcommand{\pa}{\partial}
\newcommand{\ci}{\circ}
\newcommand{\tr}{\mathrm{tr}}
\newcommand{\A}{\mathbb{A}}
\newcommand{\id}{\mathrm{id}}

\title{\bf A new approach to deformation equations \\
           of noncommutative KP hierarchies\thanks{\copyright
       2007 by A. Dimakis and F. M\"uller-Hoissen} }

\author{Aristophanes Dimakis \\
 Department of Financial and Management Engineering, \\
 University of the Aegean, 31 Fostini Str., GR-82100 Chios, Greece \\
 dimakis@aegean.gr
          \and
 Folkert M\"uller-Hoissen \\
 Max-Planck-Institute for Dynamics and Self-Organization \\
 Bunsenstrasse 10, D-37073 G\"ottingen, Germany \\
 folkert.mueller-hoissen@ds.mpg.de }

\date{}

\begin{document}

\renewcommand{\theequation} {\arabic{section}.\arabic{equation}}

\newtheorem{theorem}{Theorem}
\newtheorem{lemma}{Lemma}
\newtheorem{proposition}{Proposition}
\newtheorem{definition}{Definition}
\newtheorem{corollary}{Corollary}

\maketitle

\begin{abstract}
Partly inspired by Sato's theory of the Kadomtsev-Petviashvili (KP) hierarchy, 
we start with a quite general hierarchy of linear ordinary 
differential equations in a space of matrices and derive from it a matrix 
Riccati hierarchy. The latter is then shown to exhibit an underlying `weakly nonassociative' (WNA) algebra structure, from which we can conclude, refering 
to previous work, that any solution of the Riccati system also solves the 
potential KP hierarchy (in the corresponding matrix algebra). 
We then turn to the case where the components of the matrices are multiplied 
using a (generalized) star product. Associated with the deformation parameters, 
there are additional symmetries (flow equations) which enlarge the respective 
KP hierarchy. They have a compact formulation in terms of the WNA structure. 
We also present a formulation of the KP hierarchy equations themselves as 
deformation flow equations.
\end{abstract}

\section{Introduction}
\label{section:intro}
\setcounter{equation}{0}
This work deals with the potential Kadomtsev-Petviashvili (pKP) 
hierarchy with dependent variable $\phi$ in an associative and noncommutative 
algebra, such that the product depends on additional variables.\footnote{The 
dependent KP variable is then $u = \pa_{t_1}(\phi)$, where $t_1$ 
is the first of the infinite set of independent variables $t_1,t_2,\ldots$ 
of the KP hierarchy.}  
An important class is given by (Groenewold-) Moyal star products appearing in 
deformation quantization \cite{BFFLS78a} and, more recently, in noncommutative 
field theories \cite{Doug+Nekr01}. 
A new feature of corresponding Moyal deformed integrable equations\footnote{Here 
we mean Moyal deformations involving the independent variables which appear 
in the formulation of the respective differential equation or hierarchy. 
This has to be distinguished from the introduction of an additional `momentum' 
partner for one of these variables (see \cite{Kupe90,FMR93,Stra95,Gawr95,Stra97,Koik01,Blas+Szab03,Carr03qm}, for example), 
or an additional pair of conjugate variables, as in \cite{Taka94}. }
(see \cite{Taka01,Wang+Wada03ncsoliton,Lech+Popo04,Hama05}, 
for example) is the appearance of additional symmetries which are flows associated 
with the deformation parameters \cite{DMH00ncKdV,DMH00SW,DMH01ncNLS,Wang04} 
and which extend the corresponding hierarchies. 
The extension of the Moyal-deformed KP hierarchy has been studied in
 \cite{DMH04hier,DMH04ncKP,DMH04extMoyal,DMH05KPalgebra}. 
Moreover, \cite{DMH05KPalgebra} also dealt with certain generalized star products 
(which involve iterated Moyal-type deformations, regarding the deformation parameters 
in each step as new variables on an equal footing with the original variables 
$t_1,t_2,\ldots$). 
The present work presents another perspective on these results, which in particular 
makes them easier accessible, and moreover extends them in several ways. 
\vskip.1cm

The `deformation equations' that arise in this way are analogs of the equations 
which determine a so-called Seiberg-Witten map \cite{Seib+Witt99,Doug+Nekr01} 
in the context of `noncommutative' gauge theory. These are maps from the 
classical to the (in the sense of deformation quantization \cite{BFFLS78a}) 
`quantized' theory. 
For the Moyal-deformed (matrix) KdV equation\footnote{Here $u_t$ denotes the 
partial derivative of $u$ with respect to $t$, and $u_{xxx}$ the third 
partial derivative with respect to $x$. } 
\be
    u_t = - u_{xxx} + 3 \, ( u \ast u )_x  \, ,    \label{ncKdV}
\ee
such a `deformation flow' is given by \cite{DMH00ncKdV,DMH04hier}
\be
    u_\theta = - \frac{1}{2} [ u, u_{xx} ]_{\ast}  \, ,  \label{ncKdV_def}
\ee
where $\theta$ is the deformation parameter of the Moyal product $\ast$ in 
the space of smooth functions of $x$ and $t$, and 
$[u,v]_{\ast} := u \ast v - v \ast u$.\footnote{Appendix~A provides a FORM 
program \cite{Heck00,Verm02} to check the commutativity of the flows.} 
To every solution of the classical KdV equation, the deformation equation 
determines a solution of the deformed KdV equation \cite{DMH00ncKdV} 
(at least as a formal power series in the deformation parameter $\theta$). 
But there is more to it. The above deformation equation namely has the form 
of the (generalized) Heisenberg ferromagnet equation with `time' $\theta$, 
and it indeed reduces to it\footnote{Of course, we have to take $u$ as a 
\emph{matrix} of functions which are multiplied using the Moyal product. 
In case of vanishing deformation, the noncommutativity of the matrix product 
then remains.}
if we restrict $u$ to be independent of $t$.\footnote{The well-known relation 
with the familiar form $\vec{S}_\tau = \vec{S} \times \vec{S}_{xx}$ of 
the continuum Heisenberg ferromagnet equation (also known as Landau-Lifshitz 
equation) is obtained from $u_\theta = - \frac{1}{2} [ u, u_{xx} ]$ by choosing 
$$ u = \left( \begin{array}{cc} -S_3 & S_2 + \imath \, S_1 \\ S_2 - \imath \, S_1 & S_3
       \end{array}\right)  $$
with commuting functions satisfying $\vec{S}^2 =1$, and $\theta = \imath \, \tau$. }
The Moyal deformation thus yields an unexpected link between different 
(classical) integrable equations.\footnote{But this relation does not (at least not 
in a straight way) extend to a relation between the corresponding hierarchies,  
since the requirement that $u$ does not depend on certain $t_n$ in general leads to 
a too restrictive (and typically trivial) reduction of the (extended) KdV hierarchy. 
We should also mention that both, KdV and the Heisenberg ferromagnet equation, 
appear in branches of the AKNS system (see e.g. \cite{DMH04hier} and the references 
cited therein). Because of this reason the link between the two systems may not 
really come as a surprise. The Moyal link is, however, much more direct and 
of a very different nature. 
The AKNS framework suggests similar deformation relations between other integrable  equations, but this will not be further elaborated in this work.}
\vskip.1cm

Another example is the Boussinesq equation 
\be
    \phi_{yy} = - \frac{1}{3} \phi_{xxxx} - 2 (\phi_x \ast \phi_x)_x 
                + 2 \, [ \phi_x , \phi_y ] \, , 
\ee
where $\ast$ is now the Moyal product with respect to the variables $x$ and $y$ 
and with deformation parameter $\theta'$. In this case the corresponding 
deformation equation is the (deformed) potential KdV equation 
\be
    \phi_{\theta'} = - \frac{1}{6} \phi_{xxx} - \phi_x \ast \phi_x 
\ee
(see also section 7.2 in \cite{DMH04ncKP}), where the deformation of the 
product disappears if $\phi_y =0$. 
\vskip.1cm

These examples demonstrate that the deformation flow equations are not at all exotic 
objects, in special cases they reproduce well-known integrable systems. 
This will be further supported in section~\ref{section:Moyal} (second example) 
and in section~\ref{section:def->KP}, where deformation parameters  
are identified as the usual evolution variables of the respective integrable 
systems. 
\vskip.1cm

There is another motivation for the exploration of deformations of the 
KP hierarchy. The famous Sato theory \cite{Sato+Sato82,Taka89} expresses the 
scalar KP hierarchy as Pl\"ucker relations of an infinite-dimensional Grassmannian. 
These are algebraic identities for Pl\"ucker coordinates. A crucial step towards this geometric interpretation is to express the dependent variable of the scalar KP 
hierarchy in terms of the so-called $\tau$ function. 
In the noncommutative case, where a direct analog of the $\tau$ function is 
not available, we obtained a result of similar nature \cite{DMH05KPalgebra,DMH06nahier}: 
the (`noncommutative') pKP equations are in correspondence with a class of identities 
in the algebra of quasi-symmetric functions. It turned out, however, that there 
are further similar classes of identities. 
In order to also translate these into differential equations, one is forced to 
introduce Moyal and the abovementioned generalized deformations. The corresponding 
deformations and extensions of the pKP hierarchy thus emerge in a natural way. 
\vskip.1cm

The Sato theory \cite{Sato+Sato82,Taka89} of the scalar KP hierarchy 
(and certain generalizations) achieves to linearize it on an infinite-dimensional 
space. In this spirit section~\ref{section:linsys->KP} takes a quick step 
from a very general linear system to the noncommutative KP hierarchy. 
Here a matrix Riccati system plays a crucial role. This is further substantiated 
in section~\ref{section:WNA}, which identifies the Riccati system as a special case 
of a universal system of ordinary differential equations in a `weakly nonassociative' 
(WNA) algebra. This is based on the recent work in \cite{DMH06nahier}, see also 
\cite{DMH07Ricc,DMH07Wronski}.
After a first encounter in section~\ref{section:Moyal} with a deformation equation
of a Moyal deformed KP hierarchy, section~\ref{section:WNA-Moyal} 
offers a more systematic treatment using the WNA framework.
Section~\ref{section:beyond} then presents corresponding results for the more general
deformations (generalized star products) mentioned above. 
Section~\ref{section:def->KP} deals with a special case which puts the KP 
hierarchy in a new perspective: all evolution variables arise 
as deformation parameters! Finally, section~\ref{section:conclusions} contains 
some conclusions.

\section{From a linear system to the KP hierarchy}
\label{section:linsys->KP}
\setcounter{equation}{0}
As already mentioned in the introduction, the Sato theory \cite{Sato+Sato82,Taka89}
translates the KP hierarchy into a linear system of ordinary differential equations
(which induce commuting flows on an infinite-dimensional Grassmannian).
Let us go the inverse way and start with a linear system of ordinary differential equations\footnote{If $H$ is invertible, it may be possible to extend the system to 
$n \in \mathbb{Z}$, thus adding a `negative hierarchy'. But see also the remark 
in section~3.5.4 of \cite{Taka89} concerning possible obstructions in the 
infinite-dimensional case due to nonvanishing Kac-Peterson cocycles. }
\be
    Z_{t_n} = H^n Z  \qquad\quad  n=1,2,\ldots \, ,   \label{Z_lin_sys}
\ee
where
\be
    Z = \left(\begin{array}{c} X \\ Y \end{array} \right), \qquad
    H = \left(\begin{array}{cc} R & Q \\ S & L  \end{array} \right) \; .
             \label{Z,H-decomp}
\ee
Here $X,Y$ are $N \times N$, respectively $M \times N$ matrices with components
from an associative algebra $\mathcal{R}$, the elements of which depend smoothly
on independent variables $t_n$, $n \in \mathbb{N}$. Correspondingly, $L,Q,R,S$ are
$M \times M$, $N \times M$, $N \times N$ and $M \times N$ matrices, respectively.
Though we will concentrate on finite-dimensional matrices ($M,N \in \mathbb{N}$),
we may allow $M,N$ to be infinite (which is the case in the general
Sato theory), provided that the product of such matrices is well-defined.
The system (\ref{Z_lin_sys}) is compatible if $(H^m)_{t_n} = (H^n)_{t_m}$ for 
all $m,n$. This holds in particular if $H$ is \emph{constant}, as  
assumed in the following. 
(\ref{Z_lin_sys}) is equivalent to the linear heat hierarchy
\be
    Z_{t_n} = (\pa_{t_1})^n Z \qquad\quad  n=2,3,\ldots \, , \label{Z_heat_hier}
\ee
supplemented with
\be
    Z_{t_1} = H Z \; .
\ee
This is solved by 
\be
    Z = e^{\xi(H)} Z_0  \, , \qquad \xi(H) := \sum_{n \geq 1} t_n \, H^n \, ,
        \label{Z_sol}
\ee
where $Z_0$ is constant, i.e. independent of $t_1,t_2,\ldots$. Setting
\be
    \left(\begin{array}{cc} R_n & Q_n \\ S_n & L_n  \end{array}\right) := H^n \, ,
\ee
we obtain the recursion relations
\be
    \begin{array}{lcl}
    L_{m+n} = L_m L_n + S_m Q_n \, , & \quad &
    Q_{m+n} = R_m Q_n + Q_m L_n \, , \\
    R_{m+n} = R_m R_n + Q_m S_n \, , & \quad &
    S_{m+n} = L_m S_n + S_m R_n \, ,
    \end{array}           \label{LQRS_recurs}
\ee
where $L_1=L, Q_1=Q, R_1=R, S_1=S$. The linear system (\ref{Z_lin_sys})
decomposes into
\be
    X_{t_n} =  R_n X + Q_n Y \, , \qquad   Y_{t_n} = S_n X + L_n Y \; . 
      \label{lin_sys_XY}
\ee
\vskip.2cm

Let $\ast$ be \emph{any} (other) associative product in $\mathcal{R}$ 
which, extended to a matrix product, satisfies the relations
\be
    (L A) \ast B = L \, (A \ast B) \, , \qquad 
    (A' L) \ast B' = A' \ast (L B') \, , \qquad
    A'' \ast (B'' L) = (A'' \ast B'') \, L  \, , 
\ee
and correspondingly with $L$ replaced by $Q$, $R$ and $S$. Here  
$A,A',A'',B,B',B''$ are matrices with entries in $\mathcal{R}$ and 
appropriate dimensions, so that the matrix multiplications in these 
equations are well-defined and in particular fit to the dimensions of 
$L,Q,R,S$, respectively. 
By induction, using the recursion relations (\ref{LQRS_recurs}), one obtains  
the corresponding relations for $L_n,Q_n,R_n,S_n$, $n=2,3,\ldots$. We shall also 
assume that the partial derivatives with respect to the variables $t_n$
are derivations of the product $\ast$. 
\vskip.2cm

\begin{proposition}
For all $\mathbf{t} = (t_1,t_2,\ldots)$ for which $X$ possesses an inverse 
$X^{\ast -1}$ (with respect to $\ast$), 
\be
    \phi := Y \ast X^{\ast -1}   \label{phi_YX}
\ee
solves the pKP hierarchy in the algebra $\mathcal{A}$ of $M \times N$ matrices
with entries in $\mathcal{R}$ and product 
\be
       A \cdot B := A \ast Q B \; .   \label{astQ-product}
\ee
\end{proposition}
\textbf{Proof:} This can be proved using a functional representation of the 
pKP hierarchy, see \cite{DMH06Burgers}. An alternative proof 
will be given in this work, based on the next proposition and results 
recalled in section~\ref{section:WNA}. 
\hfill $\square$

\begin{proposition}
The linear system (\ref{Z_lin_sys}) implies the matrix Riccati system
\be
    \phi_{t_n} = S_n + L_n \phi - \phi R_n - \phi \ast Q_n \phi  \qquad n=1,2,\ldots \; .
    \label{phi_tn}
\ee 
With suitable restrictions imposed on $(\mathcal{R},\ast)$ (as specified in the proof), 
the two systems are in fact equivalent.
\end{proposition}
\textbf{Proof:} Using the definition (\ref{phi_YX}) and the derivation property 
of the partial derivatives $\pa_{t_n}$, we have
\bez
    Y_{t_n} = (\phi \ast X)_{t_n} = \phi_{t_n} \ast X + \phi \ast X_{t_n} \; .
\eez
If (\ref{Z_lin_sys}) holds, and thus (\ref{lin_sys_XY}), a direct calculation 
leads to (\ref{phi_tn}). Conversely, if (\ref{phi_tn}) holds, the above 
equation becomes
\bez
    Y_{t_n} = S_n X + L_n Y + \phi \ast ( X_{t_n} - R_n X - Q_n Y ) \, ,
\eez
and thus $Y_{t_n} - S_n X - L_n Y = Y \ast C_n$, where
$C_n := X^{\ast -1} \ast (X_{t_n} - R_n X - Q_n Y)$. Hence
\bez
       Z_{t_n} = H^n Z + Z \ast C_n \; .
\eez
The integrability conditions $Z_{t_m t_n} = Z_{t_n t_m}$ now demand that 
$C_{m,t_n} - C_{n,t_m} + [C_n , C_m]_\ast =0$, which means that 
the curvature of the connection $C = \sum_{n \geq 1} C_n \, d t_n$ vanishes. 
If this implies the existence of a gauge transformation $Z' = Z \ast G$ 
which transforms the connection to zero (as in the familiar case where 
$(\mathcal{R},\ast)$ is the algebra of functions on a chart of some manifold), 
we obtain ${Z'}_{t_n} = H^n Z'$, which is the linear system (\ref{Z_lin_sys}). 
\hfill $\square$
\vskip.2cm

Elimination\footnote{The `elimination' does \emph{not} make use of any special, 
e.g. matrix, properties of $L,R,S$, but rather treats them as abstract algebraic 
symbols. This includes a lot of freedom so that the Riccati system (\ref{phi_tn}) 
reaches large classes of solutions of the pKP hierarchy, if not even all. This 
remark applies correspondingly to the `elimination procedure' in section 
\ref{section:Moyal}. } 
of $L,R,S$ from the equations (\ref{phi_tn}) (using also equations derived from 
them by differentiations) leads to the pKP hierarchy for $\phi$
in $(\mathcal{A}, \cdot)$. A corresponding proof is given in section~\ref{section:WNA}. 
\vskip.2cm

\noindent
\textbf{Example.} In the special case where $S=0$, we have 
\be
    R_n = R^n \, , \qquad L_n = L^n \, ,
\ee
and $Q_{n+1} = Q L^n + R Q_n$. By induction, this leads to
\be
    Q_n = \sum_{k = 0}^{n-1} R^k Q L^{n-k-1} \; .  \label{Qn-S=0}
\ee
In order that $H$ (with $S=0$) can be block-diagonalized by a matrix
\be
    T =  \left(\begin{array}{cc} I_N & -K \\ 0 & I_M  \end{array}\right) \, ,
\ee
i.e. $T^{-1} H T$ is block-diagonal, we need the condition
\be
         Q = R K - K L \; .   \label{Q=RK-KL}
\ee
Then (\ref{Qn-S=0}) becomes a telescoping sum which results in
\be
    Q_n = R^n K - K L^n \; .
\ee
In this case the matrix Riccati system (\ref{phi_tn}) is solved by
\be
  \phi = e^{\xi(L)} \phi_0 \ast ( e^{\xi(R)}(I_N + K \phi_0)
         - K e^{\xi(L)} \phi_0)^{\ast -1} \, ,   \label{phi_solution}
\ee
where $\phi_0 = Y_0 \ast X_0^{\ast -1}$. Note that this solution of the 
pKP hierarchy in $(\mathcal{A},\cdot)$ is `universal' in the sense that 
we did not have to specify the associative product $\ast$. 
\vskip.1cm

(\ref{phi_tn}) with (\ref{Qn-S=0}) implies the so-called Sato system 
\cite{Taka89,FMP98} 
\be
 \mathcal{W}(i,j)_{t_n} = \mathcal{W}(i+n,j) - \mathcal{W}(i,j+n) 
    - \sum_{k=0}^{n-1} \mathcal{W}(i,k) \ast Q \, \mathcal{W}(n-k-1,j) \, , 
                \label{Sato_system}
\ee
where\footnote{If $L$ or $R$ is a finite-dimensional matrix, there are of course 
algebraic relations among the $\mathcal{W}(i,j)$ (e.g., as a consequence of the 
Cayley-Hamilton theorem). } 
\be
    \mathcal{W}(i,j) := L^i \phi R^j \qquad i,j=0,1,2,\ldots   \label{W(i,j)}
\ee
(see also \cite{DMH07Wronski}). 
If (\ref{Q=RK-KL}) holds, the system (\ref{Sato_system}) simplifies to
\be
  \mathcal{W}(i,j)_{t_n} = \mathcal{W}(i+n,j) - \mathcal{W}(i,j+n) 
    - \mathcal{W}(i,n) \ast K \, \mathcal{W}(0,j) 
    + \mathcal{W}(i,0) \ast K \, \mathcal{W}(n,j) \; .
\ee
The Sato system is well-known to be linearizable, see also the following remark.  
\hfill $\square$
\vskip.2cm

\noindent
\textbf{Remark.} If $S \neq 0$, (\ref{W(i,j)}) leads to a generalization of 
the Sato system, which can again be linearized by writing $\mathcal{W} = \mathcal{Y} \ast \mathcal{X}^{\ast -1}$. This leads to the linear system
\be
   \mathcal{Z}_{t_n} = \mathcal{H}^n \mathcal{Z} 
       \qquad \mbox{where} \quad 
    \mathcal{Z} = \left(\begin{array}{c} \mathcal{X} \\ \mathcal{Y} \end{array} \right) 
        \, , \quad 
   \mathcal{H} = \left(\begin{array}{cc}
     \Lambda^T & \mathbf{e}_0 \mathbf{e}_0^T \otimes Q \\ \mathcal{S} & \Lambda 
                       \end{array} \right)  \, , 
\ee 
with $\mathcal{S}(i,j) = L^i S R^j$, $\mathbf{e}_0^T = (1,0,\ldots)$, and 
the shift operator matrix
\be
   \Lambda = \left(\begin{array}{ccccc} 0 & 1 & 0 &   & \cdots     \\
                                        0 & 0 & 1 & 0 & \cdots     \\
                                  \vdots & \ddots & \ddots & \ddots & \ddots  
            \end{array}\right) \; .
\ee
If $S=0$, this provides a linearization of the Sato system. 
\hfill $\square$
\vskip.2cm

\noindent
\textbf{Remark.}
Let $Q = V U^T$ with a constant $M \times m$ matrix $U$ 
and a constant $N \times m$ matrix $V$, and such that $A \ast Q B = A V \ast U^T B$. 
If $\phi$ solves the pKP hierarchy in $(\mathcal{A},\cdot)$ (with the product 
(\ref{astQ-product})), then 
\be
    \varphi := U^T \phi \, V    \label{varphi}
\ee
solves the pKP hierarchy in the algebra of $m \times m$ matrices 
with product $\ast$. 

Let $\mathcal{R}$ be the algebra of smooth complex functions of 
$t_1,t_2,\ldots$, and $\ast$ the original product in $\mathcal{R}$. 
Then, for $m=1$,  
\be
    \varphi = \tr(Q \phi) 
\ee
solves the \emph{scalar} pKP hierarchy. Using 
\be
    Q \phi = X_{t_1} X^{-1} - R \, , 
\ee
which results from (\ref{lin_sys_XY}) (and reminds us of a matrix Cole-Hopf 
transformation, see also \cite{GMA94,DMH06Burgers,DMH07Ricc}), we obtain
\be
    \varphi = \tr(X_{t_1} X^{-1}) - \tr(R) = (\log\tau)_{t_1} - \tr(R)
\ee
with
\be
    \tau = \det(X) \; .
\ee
In particular, any solution of the matrix Riccati system (\ref{phi_tn}) with
$\mathrm{rank}(Q)=1$ determines in this way a $\tau$ function (and then a solution)
of the scalar KP hierarchy. 
The solution (\ref{phi_solution}) of the matrix pKP hierarchy leads in this 
way to a large set of solutions of the scalar KP hierarchy (see also \cite{DMH06Burgers,DMH06CJP,DMH07Ricc} and the references therein).
\hfill $\square$

\section{A nonassociative algebraic structure underlying the Riccati system}
\label{section:WNA}
\setcounter{equation}{0}
Let $(\mathcal{A},\cdot)$ denote the algebra of $M \times N$ matrices
considered in the preceding section. In this algebra we introduce the 
sequence of products
\be
    A \ci_n B = A \ast Q_n B \qquad \quad n=1,2, \ldots 
\ee
(so that $A \ci_1 B = A \cdot B$), which are combined associative, i.e.,
\be
    A \ci_n (B \ci_m C) = (A \ci_n B) \ci_m C \; .
\ee
We augment the algebra $\mathcal{A}$ with an element $\nu$ such that
\be
    \nu \ci_n \nu = - S_n \, , \qquad
    \nu \ci_n A = L_n A \, , \qquad
    A \ci_n \nu = - A R_n \; .
\ee
Let $\A$ denote the resulting algebra. From the recursion relations 
(\ref{LQRS_recurs}) we obtain
\be
    a \ci_{m+n} b = a \ci_m (\nu \ci_n b) - (a \ci_m \nu) \ci_n b  
     \label{ci_m+n}
\ee
for all $a,b \in \A$. This relation determines all the products $\ci_n$ in terms 
of $\ci_1$. 
If $a \ci_2 b \neq 0$ for some $a,b \in \A$, then $\A$ is \emph{not} associative. 
Nonassociativity only enters through the augmented element $\nu$. 
This motivates the following definition \cite{DMH06nahier} (see also \cite{DMH07Ricc}).
\vskip.2cm

An algebra $(\A,\circ)$ (over a commutative ring) is called \emph{weakly
nonassociative (WNA)} if it is not associative, but the associator 
$(a,b \circ c,d)$ vanishes for all $a,b,c,d \in \A$ (where the associator is 
defined as $(a,b,c) = (a \circ b) \circ c - a \circ (b \circ c)$).
The middle nucleus
$\A' = \{ b \in \A  \, | \, (a,b,c) = 0 \; \forall a,c \in \A \}$, which is an
\emph{associative} subalgebra, is then also an ideal in $\A$.
With respect to an element $f \in \A$, $f \not\in \A'$, 
we define the products $a \circ_1 b = a \circ b$ and
\be
        a \circ_{n+1} b = a \circ (f \circ_n b) - (a \circ f) \circ_n b
        \qquad \quad  n=1,2, \ldots \; .    \label{circ_n}
\ee
As a consequence of the WNA condition, these products only depend on the equivalence
class $[f]$ of $f$ in the quotient space $\A/\A'$. Next we recall a central 
result from \cite{DMH06nahier}.

\begin{theorem}
\label{theorem:WNA}
Let $(\A,\circ)$ be any WNA algebra, the elements of which depend smoothly
on independent variables $t_1,t_2, \ldots$.
Then the flows of the system of ordinary differential equations
\be
    f_{t_n} = f \circ_n f  \qquad \quad   n=1,2, \ldots   \label{f_tn}
\ee
in $\A$ commute and imply that $-f_{t_1} \in \A'$ solves the KP hierarchy 
in the associative subalgebra $(\A',\circ)$.
\hfill $\square$
\end{theorem}
\vskip.2cm

The algebra $\A$ introduced above is easily seen to be WNA. Setting
\be
     f = \nu - \phi \, ,  \label{f=nu-phi}
\ee
so that $[f]=[\nu]$, then (\ref{circ_n}) coincides with 
(\ref{ci_m+n}) for $m=1$. Furthermore, (\ref{circ_n}) implies (\ref{ci_m+n}) 
(see proposition~3.3 in \cite{DMH06nahier}).  
Assuming that $\nu$ is constant (i.e., independent of
$t_1,t_2,\ldots$), (\ref{phi_tn}) becomes (\ref{f_tn}). 
By application of the theorem, it follows that if $\phi$ solves the 
Riccati system (\ref{phi_tn}), then it also solves the pKP hierarchy in 
$(\mathcal{A},\cdot)$. 
\vskip.2cm

Of course, there are other realizations of WNA algebras than given by 
the class of examples which we encountered in this section, in particular 
a realization in terms of quasi-symmetric functions \cite{DMH06nahier} and 
a realization in terms of formal pseudodifferential operators \cite{DMH07Ricc} 
(which allows to make contact with the Gelfand-Dickey formalism \cite{Dick03} 
of the KP hierarchy).

\section{The case of a Moyal star product}
\label{section:Moyal}
\setcounter{equation}{0}
New structures appear if the product $\ast$ depends on additional variables. 
Of particular interest is the case of a (Groenewold-) Moyal star product
\be
  g_1 \ast g_2 := \mathbf{m} \; \exp\Big( \frac{1}{2} \sum_{m,n \geq 1} \theta_{mn}
        \, \pa_{t_m} \otimes \pa_{t_n} \Big) \; g_1 \otimes g_2  
       \label{Moyal-prod}
\ee
with antisymmetric (deformation) parameters $\theta_{mn}$ and
$\mathbf{m}(g_1 \otimes g_2) = g_1 \, g_2$.\footnote{The tensor product 
has to be taken over $\mathbb{K}[[\{\theta_{mn}\}]]$ where $\mathbb{K}$ is 
$\mathbb{R}$ or $\mathbb{C}$.} 
The product extends to matrices by combining the ordinary matrix product 
with the Moyal product of the components. 
So far this just gives a class of examples of noncommutative associative products 
to which our results in the previous sections apply. 
But now we can think of the deformation parameters as new variables which 
describe deformation flows of the deformed KP hierarchy. Since 
\be
   (g_1 \ast g_2)_{\theta_{mn}}
 =  g_{1,\theta_{mn}} \ast g_2 + g_1 \ast g_{2,\theta_{mn}}
    + \frac{1}{2} (g_{1,t_m} \ast g_{2,t_n} - g_{1,t_n} \ast g_{2,t_m}) \, ,
\ee
there is a non-trivial coupling of these flows to those of the pKP 
hierarchy.\footnote{If $\tilde{\mathbf{m}}$ denotes any product which depends on a 
parameter $\theta$, then  
$\pa_\theta ( \tilde{\mathbf{m}}(g_1,g_2)) = \tilde{\mathbf{m}}(\pa_\theta(g_1), g_2) 
 + \tilde{\mathbf{m}}(g_1,\pa_\theta(g_2)) + \pa_\theta(\tilde{\mathbf{m}})(g_1,g_2)$. 
In case of the Moyal product, we have
$\pa_{\theta_{mn}}(\tilde{\mathbf{m}})(g_1,g_2) = \frac{1}{2} ( \tilde{\mathbf{m}}(g_{1,t_m}, g_{2,t_n}) 
  - \tilde{\mathbf{m}}( g_{1,t_n}, g_{2,t_m}) )$, so that
$\pa_{\theta_{mn}}(\tilde{\mathbf{m}})$ 
can be expressed in terms of $\tilde{\mathbf{m}}$. This property is crucial for the 
following calculations. }
By considering deformation flows we are actually leaving the framework of 
the KP hierarchy in a fixed algebra (i.e., with a fixed product). Instead we 
are dealing with a \emph{family} of KP hierarchies, parametrized by 
the deformation parameters entering the product. This point of view is 
supported by the fact that the KP hierarchy possesses solutions which can 
be expressed without the need of specifying the concrete form of the product, 
see the example in section~\ref{section:linsys->KP}. In the following we will 
not be so careful to distinguish between a deformed KP hierarchy and the 
family of deformed KP hierarchies. It is the latter which we are studying. 
\vskip.2cm

Let us try to extend the linear system (\ref{Z_lin_sys}) by supplementing it 
with compatible linear ordinary differential equations in the variables 
$\theta_{mn}$. Using only what we already have at our disposal, we should 
build the corresponding vector field in terms of $H$. But because of the 
antisymmetry of the variables $\theta_{mn}$, we are left with the apparently 
trivial choice 
\be
    Z_{\theta_{mn}} = 0  \qquad m,n=1,2,\ldots \; . 
\ee
Nevertheless, this leads to non-trivial results since now (\ref{phi_YX}) 
involves the deformation parameters through the product. In fact, we obtain
\be
    0 = Y_{\theta_{mn}} = (\phi \ast X)_{\theta_{mn}} 
      = \phi_{\theta_{mn}} \ast X + {1\over 2}(\phi_{t_m} \ast X_{t_n} 
        - \phi_{t_n} \ast X_{t_m}) \, ,
\ee
and thus
\be
   \phi_{\theta_{mn}} = \frac{1}{2} ( \phi_{t_n} \ast X_{t_m} 
       - \phi_{t_m} \ast X_{t_n}) \ast X^{\ast -1}  \; . \label{pre_phi_thmn}
\ee
By use of (\ref{lin_sys_XY}), this implies the deformation flow equations
\be
    \phi_{\theta_{mn}} 
  = \frac{1}{2} \Big(\phi_{t_n} \ast (R_m + Q_m \phi)
                - \phi_{t_m} \ast (R_n + Q_n \phi) \Big)  \label{phi_thmn} \; .
\ee
With the help of the Riccati system (\ref{phi_tn}), these are converted into 
the \emph{ordinary} differential equations 
\be
     \phi_{\theta_{mn}} 
 &=& \frac{1}{2} \Big( S_n R_m - S_m R_n + (S_n Q_m - S_m Q_n) \, \phi 
     + L_n \phi R_m - L_m \phi R_n - \phi \, [R_n , R_m] \nonumber \\
 & & + L_n \phi \ast Q_m \phi - L_m \phi \ast Q_n \phi 
     - \phi \ast (R_n Q_m - R_m Q_n) \phi 
     - \phi \ast Q_n \phi R_m + \phi \ast Q_m \phi R_n \nonumber \\
 & & - \phi \ast Q_n \phi \ast Q_m \phi 
     + \phi \ast Q_m \phi \ast Q_n \phi \Big)
       \qquad\qquad  m,n=1,2,\ldots \; .   \label{phi_thmn_ode}
\ee
Because of the commutativity of flows on the level of the linear system, 
we should expect these deformation flows to commute with each other, 
and also with the flows of the Riccati system (\ref{phi_tn}). 
There is a subtlety, however, which arises from the elimination of the 
terms $X_{t_n} \ast X^{\ast -1}$ in the step from (\ref{pre_phi_thmn}) 
to (\ref{phi_thmn}). 
Whereas any of the flows (\ref{pre_phi_thmn}) indeed commutes with any 
of the flows of the Riccati system (\ref{phi_tn}) without further conditions, 
the commutativity of (\ref{phi_thmn}) or (\ref{phi_thmn_ode}) for fixed $m,n$ 
with the $t_r$-flow of (\ref{phi_tn}) also requires the two equations of 
(\ref{phi_tn}) with evolution variables $t_m$ and $t_n$. We will refer 
to this feature as `conditional commutativity'.
\vskip.2cm

Eliminating $L,R,S$ (regarded as abstract algebraic symbols) from the above  
equations, results in deformation equations for $\phi$, the simplest of which is
\be
    \phi_{\theta_{12}} = \frac{1}{6} (\phi_{t_3} - \phi_{t_1 t_1 t_1})
                         - \phi_{t_1} \ast Q \phi_{t_1} \; .  \label{phi_th12}
\ee
This already requires a lengthy calculation (see also appendix~B), which asks
for a method to achieve such results in a more systematic and efficient way,
a problem addressed in the following section. The deformed pKP hierarchy 
together with its deformation equations will be called `extended (deformed) 
pKP hierarchy'. The deformation flows commute with each other and with 
those of the pKP hierarchy conditionally in the sense explained above. 
\vskip.2cm

An equation like (\ref{phi_th12}) is \emph{not} a symmetry of the pKP hierarchy 
in a given algebra (which would mean fixing the deformation parameters in 
the product (\ref{Moyal-prod})), but rather a symmetry of the \emph{family}  
of pKP hierarchies, which is parametrized in terms of the deformation parameters. 
The corresponding flow maps solutions of one pKP hierarchy to solutions 
of another pKP hierarchy (with different values of the parameters $\theta_{mn}$). 
\vskip.2cm

\noindent
\textbf{Remark.} If $S=0$ and $Q = R K - K L$, we have the solution 
(\ref{phi_solution}) of the Moyal-deformed matrix Riccati system and thus 
the Moyal-deformed matrix pKP hierarchy (with matrix product modified by $Q$). 
It also solves all the deformation equations, like (\ref{phi_th12}). 
\hfill $\square$
\vskip.2cm

\noindent
\textbf{Remark.}
If $\phi$ solves the extended deformed pKP hierarchy in the matrix algebra 
with product $A \cdot B = A \ast Q B$, then 
(\ref{varphi}) solves the extended deformed pKP hierarchy in the algebra 
of $m \times m$ matrices with entries in the Moyal algebra. 
(\ref{phi_th12}) then leads to
\be
    \varphi_{\theta_{12}} = \frac{1}{6} (\varphi_{t_3} - \varphi_{t_1 t_1 t_1}) 
      - \varphi_{t_1} \ast \varphi_{t_1} \; . 
\ee
This equation (for $m=1$) first appeared in \cite{DMH04hier}, 
see also \cite{DMH04ncKP,DMH04extMoyal,DMH05KPalgebra}. 
\hfill $\square$
\vskip.2cm

\noindent
\textbf{Example.} We show that, in a special case, a deformation flow 
equation reproduces the potential KdV equation such that the deformation 
parameter takes the role of its (usual) time variable. 
Let us consider only the first two equations of the linear system, i.e. 
\be
    Z_x = H Z \, , \qquad Z_y = H^2 Z \, ,  \label{ex_linsys}
\ee
where $x=t_1$ and $y=t_2$. Assuming moreover 
\be
     \phi_y=0 \, , 
\ee
this leads to 
\be
    \phi_x = S + L \phi - \phi R - \phi Q \phi \, , \qquad
    S_2 + L_2 \phi - \phi R_2 - \phi Q_2 \phi = 0 \, ,  \label{ex_algRicc}
\ee
which is not affected by the deformation so that only the original product appears.  
 From $Z_\theta =0$, where $\theta=\theta_{12}$, we find
\be
    0 = Y_\theta = \phi_\theta \ast X + \frac{1}{2} \phi_x \ast X_y \, , 
\ee
and thus, by use of (\ref{ex_linsys}) (decomposed as in (\ref{lin_sys_XY})), 
\be
    \phi_\theta = - \frac{1}{2} \phi_x \, (R_2 + Q_2 \phi) \; . \label{ex_phi_theta}
\ee
A straightforward but tedious calculation, eliminating $L,R,S$ from our three 
equations for $\phi$, leads to\footnote{In order to check that this equation 
holds, one first eliminates the derivatives of $\phi$ by use of our previous equations 
for $\phi_x$ and $\phi_\theta$. The resulting algebraic equation is then 
satisfied as a consequence of the algebraic Riccati equation in (\ref{ex_algRicc}) 
which resulted from setting $\phi_y=0$. This is easily verified using FORM.}
\be
   \phi_\theta = - \frac{1}{8} \phi_{xxx} - \frac{3}{4} \phi_x Q \phi_x \; .
                 \label{ex_KdV}
\ee
This is the \emph{potential KdV equation}, where the deformation parameter 
$\theta$ plays the role of the evolution `time' variable $t$.\footnote{After 
suitable rescalings of the variables, it takes the form (\ref{ncKdV}).}
Here the KdV flow originates from a Moyal deformation! See also section 7.1 in \cite{DMH04ncKP} for related results. 
 From (\ref{ex_linsys}) and $\phi_y=0$, we obtain $Y_y = \phi \ast X_y 
= Y \ast C$ with an $N \times N$ matrix $C$ such that 
$R_2 X + Q_2 Y = X \ast C$. Furthermore, we have 
$S_2 X + L_2 Y = Y \ast C$, which combines with the last equation to
\be
    H^2 Z = Z \ast C \, , 
\ee
where on the right hand side both parts ($X$ and $Y$) of $Z$ are multiplied by $C$.
Since the solution of the above linear system is given by
\be
    Z = e^{x H + y H^2} Z_0 \, , 
\ee
it follows that
\be
    H^2 Z_0 = Z_0 \, C \, ,
\ee
which is a condition on the initial data given by $Z_0$. Conversely, this 
condition implies $\phi_y=0$ by use of (\ref{ex_linsys}). 
In conclusion, any solution of the linear system (\ref{ex_linsys}) 
and $Z_\theta =0$, with initial data satisfying $H^2 Z_0 = Z_0 C$ with 
some $N \times N$ matrix $C$, solves the potential KdV equation (\ref{ex_KdV}).
\hfill $\square$

\section{The WNA version of the Moyal deformation equations}
\label{section:WNA-Moyal}
\setcounter{equation}{0}
Using (\ref{f=nu-phi}) and the WNA rules of section~\ref{section:WNA}, 
where now $\ast$ is chosen to be the Moyal product defined in the preceding section, 
the system (\ref{phi_thmn}) can be expressed in the form
\be
    f_{\theta_{mn}} = \frac{1}{2} ( f_{t_m} \ci_n f - f_{t_n} \ci_m f)
  = \frac{1}{2} \Big( (f \ci_m f) \ci_n f - (f \ci_n f) \ci_m f \Big)
    \qquad m,n=1,2,\ldots \, , 
\ee
by use of (\ref{f_tn}). With the help of (\ref{ci_m+n}) this can be rewritten as
\be
  f_{\theta_{mn}} = \frac{1}{2} \Big(f \ci_m (f \ci_n f) - f \ci_n (f \ci_m f) \Big)
    \; .  \label{f_thmn}
\ee
Next we address the commutativity of these flows, and also with those 
given by (\ref{f_tn}), by solely using general 
properties of WNA algebras. With this step we proceed beyond the special 
realization of a WNA algebra given in section~\ref{section:WNA}.

\begin{lemma}
Let $\A$ be a Moyal-deformed WNA algebra and $f \in \A$ such that 
$\pa_{t_n}(f) \in \A'$ and $\pa_{\theta_{mn}}(f) \in \A'$ for all $m,n=1,2,\ldots$. 
Then, for $k=1,2,\ldots$, we have
\be
    \pa_{\theta_{mn}}(a \ci_k b) = \pa_{\theta_{mn}}(a) \ci_k b 
  + a \ci_k \pa_{\theta_{mn}}(b) + \frac{1}{2} \Big( \pa_{t_m}(a) \ci_k \pa_{t_n}(b) 
    - \pa_{t_n}(a) \ci_k \pa_{t_m}(b) \Big) \; . 
\ee
\end{lemma}
{\em Proof:} by induction on $k$, using (\ref{circ_n}). 
\hfill $\square$

\begin{proposition}
\label{prop:Moyal_comm}
For any Moyal-deformed WNA algebra, the flows (\ref{f_thmn}) commute with 
the flows (\ref{f_tn}) and with each other.\footnote{More precisely, 
this is a `conditional commutativity' as discussed in section~\ref{section:Moyal}. 
The commutativity of the two flows with evolution variables $t_{mn}$ and $t_{rs}$ 
requires also the four equations of (\ref{f_tn}) with evolution variables 
$t_m,t_n,t_r,t_s$. } 
\end{proposition}
{\em Proof:} As a consequence of (\ref{f_tn}) and (\ref{f_thmn}), 
the assumptions of the preceding lemma are fulfilled. Hence
\bez
    (f_{t_r})_{\theta_{mn}} = (f \ci_r f)_{\theta_{mn}} 
  = f_{\theta_{mn}} \ci_r f + f \ci_r f_{\theta_{mn}} 
    + \frac{1}{2} ( f_{t_m} \ci_r f_{t_n} - f_{t_n} \ci_r f_{t_m} ) \, ,
\eez
which has to be further worked out with the help of (\ref{f_thmn}), and 
(\ref{f_tn}) for $m$ and $n$. A similar calculation evaluates 
$(f_{\theta_{mn}})_{t_r}$. Taking the difference of the results leads to
\bez
     (f_{t_r})_{\theta_{mn}} - (f_{\theta_{mn}})_{t_r}
  = \frac{1}{2} ( F_{mnr} - F_{nmr} ) 
\eez
with
\bez
      F_{mnr} 
 &:=& f \ci_m ((f \ci_n f) \ci_r f) - f \ci_m (f \ci_n (f \ci_r f)) 
     + f \ci_r (f \ci_m (f \ci_n f)) \\
 & & - (f \ci_r f) \ci_m (f \ci_n f) + (f \ci_m f) \ci_r (f \ci_n f) 
     - f \ci_m ( (f \ci_r f) \ci_n f) \, ,
\eez
which turns out to vanish by application of proposition~3.3 in \cite{DMH06nahier} 
(which is (\ref{ci_m+n}) with $\nu$ replaced by $f$). 
The commutativity of two deformation flows, under the condition that the 
associated equations of the hierarchy (\ref{f_tn}) are satisfied, is verified 
in the same way with little more efforts. We omit the details since this result 
can also be deduced from the more general proposition~\ref{prop:genhier_comm} 
below (since $f_{\theta_{mn}} = (f_{t_{mn}}-f_{t_{nm}})/2$). 
\hfill $\square$
\vskip.2cm

In the preceding section we were led to the problem of eliminating 
$L,R,S$ from the deformation flow equations (\ref{phi_thmn_ode}), 
in order to obtain partial differential equations involving $\phi$ (and $Q$) 
only. Since according to section~\ref{section:WNA} they are encoded 
in the action of the constant augmented element $\nu = f + \phi$, this means we have to look for equations resulting from (\ref{f_thmn}) which 
do not contain a `bare' $f$, i.e. an $f$ without a partial derivative acting 
on it (note that $f_{t_n} = -\phi_{t_n}$).\footnote{In particular, these 
restrictions rule out expressions involving 
`higher' products $\ci_n$, $n > 1$, of elements of $\A'$, since they contain  
bare $f$'s, see the definition (\ref{circ_n}).} 
Unfortunately there is no simple way to extract such equations from (\ref{f_thmn}).  
The next result is a generalization of Lemma~5.1 in \cite{DMH06nahier}. 
Recall that $\ci$ and $\ci_1$ denote the same product.

\begin{lemma}
In a WNA algebra with the products $\ci_n$ defined in (\ref{circ_n}), the 
following identities hold,
\be
  f \ci_n a &=&  L_f^n a - \sum_{k=1}^{n-1} p_{n-k} \ci L_f^{k-1} a \, ,
             \label{fci_na} \\
  a \ci_n f &=& (-1)^{n+1} R_f^n a - \sum_{k=1}^{n-1}(-1)^k (R_f^{k-1}a) \ci p_{n-k}
    \; .
\ee
where
\be
         p_n := f \ci_n f \, ,
\ee
and $L_f,R_f$ denote, respectively, left and right action by $f$.
\end{lemma}
{\em Proof:} By use of proposition~3.3 in \cite{DMH06nahier} (which is (\ref{ci_m+n}) 
with $\nu$ replaced by $f$) we have
\bez
    \sum_{k=1}^{n-1} p_{n-k} \ci L_f^{k-1} a
 = \sum_{k=1}^{n-1} (f \ci_{n-k} f) \ci_1 L_f^{k-1} a
 = \sum_{k=1}^{n-1} \Big( f \ci_{n-k} (f \ci_1 L_f^{k-1} a)
    - f \ci_{n-k+1} L_f^{k-1} a \Big) \, ,
\eez
which is a telescoping sum that collapses to $L_f^n a - f \ci_n a$. This proves
the first assertion, and the second is verified in the same way.
\hfill $\square$
\vskip.2cm

 As an application of this lemma, we find
\be
    f \ci_m (f \ci_n f) = h^{(n)}_m - \sum_{k=1}^{m-1} p_{m-k} \ci h^{(n)}_{k-1} \, ,
        \label{fm(fnf)}
\ee
where we introduced
\be
     h^{(n)}_m = L_f^m p_n  \qquad m=0,1,\ldots, \quad n=1,2,\ldots \; .
                          \label{h^(n)_m}
\ee
Note that
\be
    h^{(1)}_m = L_f^{m+1}f =: h_{m+1} \, , \qquad h^{(n)}_0 = p_n \; .
\ee
With the help of (\ref{fm(fnf)}) we can now rewrite (\ref{f_thmn}). 
Next we look for efficient formulae to compute the expressions (\ref{h^(n)_m}).

\begin{proposition}
As a consequence of (\ref{f_tn}) we have\footnote{For $r=1$ this becomes
$h^{(n+1)}_m = h^{(n)}_{m+1} + h_{m+n+1} -\pa_{t_1} \ci h^{(n)}_m
 + \sum_{k=1}^m h_k \ci h^{(n)}_{m-k} - \sum_{k=1}^{n-1} h^{(n-k)}_m \ci h_k$,
a relation which already appeared, with different notation, as (5.16) in 
\cite{DMH04ncKP} and as (4.14) in \cite{DMH05KPalgebra}.}
\be
    h^{(r+n)}_m = h^{(n)}_{r+m} + h^{(r)}_{m+n} - \pa_{t_r}\,h^{(n)}_m
        + \sum_{k=1}^m h^{(r)}_{k-1} \ci h^{(n)}_{m-k}
        - \sum_{k=1}^{r-1} h^{(r-k)}_m \ci h^{(n)}_{k-1}
        - \sum_{k=1}^{n-1} h^{(n-k)}_m \ci h^{(r)}_{k-1} \; .  \label{h^(n)_m_recurs}
\ee
\end{proposition}
{\em Proof:} First we note that
\bez
     \pa_{t_r} h_1^{(n)}
   = f_{t_r} \ci (f \ci_n f) + f \ci ( f_{t_r} \ci_n f + f \ci_n f_{t_r} )
   = p_r \ci p_n + f \ci (p_r \ci_n f + f \ci_n p_r) \; .
\eez
With the help of $f \ci_{r+n} f = f \ci_r p_n - p_r \ci_n f$ (proposition~3.3 
in \cite{DMH06nahier}), this is recognized as the $m=1$ case of
\bez
    \pa_{t_r} h^{(n)}_m = \sum_{k=1}^m h^{(r)}_{k-1} \ci h^{(n)}_{m-k} +
            L_f^m\Big(f\ci_r(f\ci_n f) + f\ci_n(f\ci_r f)\Big) - h^{(r+n)}_m \, ,
\eez
which is then easily proved by induction on $m$ using
$\pa_{t_r} h^{(n)}_{m+1} = \pa_{t_r} ( f \ci h^{(n)}_m ) = p_r \ci h^{(n)}_m
+ f \ci \pa_{t_r} h^{(n)}_m$.
Finally we make use of (\ref{fm(fnf)}) to translate the last formula into  (\ref{h^(n)_m_recurs}).
\hfill $\square$
\vskip.2cm

 For $n=1$, (\ref{h^(n)_m_recurs}) reduces to
\be
   h^{(r+1)}_m = h_{m+r+1} + h^{(r)}_{m+1} - \pa_{t_r}\,h_{m+1}
 + \sum_{k=1}^m h^{(r)}_{k-1} \ci h_{m-k+1} 
 - \sum_{k=1}^{r-1} h^{(r-k)}_m \ci h_k \, ,    \label{h^(r+1)_m_recurs}
\ee
which allows a recursive computation of the $h^{(n)}_m$ in terms of the
$h_k$ and their derivatives. In particular, from (\ref{fm(fnf)}) and (\ref{h^(r+1)_m_recurs}) we obtain
\be
    f \ci_1 (f \ci_2 f) &=& h^{(2)}_1 = 2 h_3 - (h_2)_{t_1} + p_1 \ci p_1 \, ,  \\
    f \ci_2 (f \ci_1 f) &=& h^{(1)}_2 - p_1 \ci h^{(1)}_0 = h_3 - p_1 \ci p_1 \, ,
\ee
so that
\be
    f \ci_1 (f \ci_2 f) - f \ci_2 (f \ci_1 f) = h_3 - (h_2)_{t_1} + 2 p_1 \ci p_1 \; .
\ee
Theorem~5.3 in \cite{DMH06nahier}, which makes use of the 
hierarchy equations (\ref{f_tn}), expresses $h_n$ as 
\be
           h_n = \mathbf{p}_n(\tilde{\pa})\, f 
\ee
in terms of the elementary Schur polynomial $\mathbf{p}_n$ and 
$\tilde{\pa} := (\pa_{t_1}, \pa_{t_2}/2, \pa_{t_3}/3, \ldots)$. 
In particular, we find
\be
    h_2 = \frac{1}{2} f_{t_2} + \frac{1}{2} f_{t_1 t_1} \, , \qquad
    h_3 = \frac{1}{3} f_{t_3} + \frac{1}{2} f_{t_2 t_1} + \frac{1}{6} f_{t_1 t_1 t_1} \; .
\ee
Thus we obtain
\be
    f_{\theta_{12}} = \frac{1}{6} ( f_{t_3} - f_{t_1 t_1 t_1} ) + f_{t_1} \ci f_{t_1} \, ,
\ee
which, by use of (\ref{f=nu-phi}), reproduces our previous equation (\ref{phi_th12}).
The advantage is that now we have a systematic way to derive such deformation equations. This method is considerably simpler than that described in \cite{DMH05KPalgebra}.

\section{Beyond Moyal deformation}
\label{section:beyond}
\setcounter{equation}{0}
Instead of the Moyal deformation, let us consider the deformation\footnote{The 
tensor product has to be taken over $\mathbb{K}[[\{t_{mn}\}]]$.} 
\be
    g_1 \ast_2 g_2 := \mathbf{m}_2( g_1 \otimes g_2 )
    := \mathbf{m}\; \exp\Big(\sum_{m,n \geq 1} t_{mn} \;\pa_{t_m}
                    \otimes \pa_{t_n}\Big) \, g_1 \otimes g_2 \, ,
\ee
where no restriction is imposed on the parameters $t_{mn}$. 
The product $\ast_2$ may be regarded as being composed of
Moyal deformations with parameters $\theta_{mn} = (t_{mn}-t_{nm})/2$ and 
`trivial' deformations associated with the symmetric parts of the $t_{mn}$. 
The latter can be eliminated by means of an equivalence transformation 
(see \cite{Gutt+Rawn99}, for example, and also \cite{DMH05KPalgebra}).\footnote{A 
deformation with asymmetric $t_{mn}$ (which is equivalent to a Moyal deformation) 
appeared as follows in the context of integrable systems. The operation of taking 
the symbol $\mathrm{sym}(P) := \sum_{-\infty < j \ll \infty} a_j \, p^j$ 
of a pseudo-differential operator $P = \sum_{-\infty < j \ll \infty} a_j \, \pa_x^j$ 
has the homomorphism property 
$\mathrm{sym}(P P') = \mathrm{sym}(P) \ast_{\mathrm{sym}} \mathrm{sym}(P')$, 
where $a \ast_{\mathrm{sym}} b = \mathbf{m} \; \exp( \pa_x \otimes \pa_p ) 
\, a \otimes b$ (see \cite{FMR93,Stra97}, for example). This allows to 
translate the Gelfand-Dickey formalism \cite{Dick03} of integrable systems 
into star product language. This is not, however, of relevance for the present work. } 
We have the generalized Leibniz rule
\be
  \pa_{t_{mn}}(g_1 \ast_2 g_2) = (\pa_{t_{mn}}g_1)\ast_2 g_2
     + (\pa_{t_m} g_1)\ast_2 (\pa_{t_n} g_2) + g_1 \ast_2 (\pa_{t_{mn}}g_2) \; .
\ee
Now we proceed as in section~\ref{section:Moyal}.

\begin{proposition}
\label{prop:Ztmn-equiv}
By use of (\ref{Z_lin_sys}), $Z_{t_{mn}}=0$ implies  
\be
   \phi_{t_{mn}} = - \phi_{t_m} \ast_2 (R_n + Q_n \phi) \; .  \label{phi_tmn}
\ee
Under certain conditions (see the proof) also the converse holds. 
\end{proposition}
\textbf{Proof:} In the following $\ast$ stands for $\ast_2$. 
Using $X_{t_n} = R_n X + Q_n Y$, we find
\bez
     Y_{t_{mn}} 
 &=& (\phi \ast X)_{t_{mn}} 
  = \phi_{t_{mn}} \ast X + \phi_{t_m} \ast X_{t_n} + \phi \ast X_{t_{mn}} \\
 &=& ( \phi_{t_{mn}} + \phi_{t_m} \ast (R_n + Q_n \phi) ) \ast X 
     + \phi \ast X_{t_{mn}} 
\eez
If $Z_{t_{mn}}=0$, this implies (\ref{phi_tmn}). Conversely, if (\ref{phi_tmn}) 
holds, then we have $Y_{t_{mn}} = Y \ast C_{mn}$ with 
$C_{mn} = X^{\ast -1} \ast X_{t_{mn}}$, hence $Z_{t_{mn}} = Z \ast C_{mn}$. 
By use of (\ref{Z_lin_sys}), the integrability conditions 
$Z_{t_{mn} t_r} =Z_{t_r t_{mn}}$ lead to $C_{mn,t_r} =0$. 
The remaining integrability conditions $Z_{t_{mn} t_{rs}} = Z_{t_{rs} t_{mn}}$ 
then state that the connection $C := \sum_{m,n \geq 1} C_{mn} \, d t_{mn}$ 
has vanishing curvature. Provided that this implies the existence of a 
gauge transformation $Z' = Z \ast G$ such that the transformed connection 
vanishes, we obtain ${Z'}_{t_{mn}} = 0$. 
\hfill $\square$
\vskip.2cm

Translating (\ref{phi_tmn}) into the WNA language, using the rules of 
section~\ref{section:WNA} and $A \ci_1 B = A \ast_2 QB$, we find the 
following deformation equations
\be
    f_{t_{mn}} = f_{t_m} \ci_n f  \qquad \quad m,n=1,2,\ldots \; .
\ee
By use of $f_{t_m} = f \ci_m f$, this becomes a system of ordinary 
differential equations,
\be
     f_{t_{mn}}
  = (f \ci_m f) \ci_n f = f \ci_m (f \ci_n f) - f \ci_{m+n} f \; .
          \label{f_tmn}
\ee
The commutativity condition for an equation 
of the system (\ref{f_tmn}), with fixed $m$ and $n$, and $f_{t_r} = f \ci_r f$ 
also requires that $f_{t_m} = f \ci_m f$ and $f_{t_n} = f \ci_n f$ are satisfied. 
Furthermore, in order that two flows of the system (\ref{f_tmn}) commute, also 
the associated four equations of the hierarchy (\ref{f_tn}) are needed. 
The origin of this conditional commutativity has already been identified 
in section~\ref{section:Moyal}. 
\vskip.2cm

\noindent
\textbf{Remark.} Assuming 
$Z_{t_{mn}}= H^{m+n} Z$ instead of $Z_{t_{mn}}=0$, a straightforward calculation 
leads to $f_{t_{mn}} = f \ci_m (f \ci_n f)$ instead of (\ref{f_tmn}). 
The origin of this alternative 
can be understood as follows. In general, $Z_0$ in the solution (\ref{Z_sol}) 
to the linear system (\ref{Z_lin_sys}) may depend on the deformation parameters. 
Choosing the solution $Z' = e^{t_1H} \ast_2 e^{t_2H^2} \ast_2 \cdots Z_0$, where 
$(Z_0)_{t_{mn}} =0$, this can be evaluated to $Z' = e^{\xi(H)} Z'_0$, where the 
dependence on deformation parameters is absorbed in $Z'_0$. But now 
$Z'$ solves $Z_{t_{mn}}= H^{m+n} Z$. The second equality of (\ref{f_tmn}) shows 
that the new deformation equation differs from the old by the rhs of an equation 
of the hierarchy (\ref{f_tn}). 
\hfill $\square$
\vskip.2cm

We can proceed to a further deformation by setting\footnote{Here the tensor 
product has to be taken over $\mathbb{K}[[\{t_{mnr}\}]]$. 
But before application of $\mathbf{m}_2$, it has to be projected to 
a tensor product over $\mathbb{K}[[\{t_{kl},t_{mnr}\}]]$ (and $\mathbf{m}_2$ 
then has to be built with the latter). 
Our notation is a bit sloppy in this respect. But this should not really 
cause problems in actual calculations. \label{foot:tensor-prod} }
\be
    g_1 \ast_3 g_2 := \mathbf{m}_3( g_1 \otimes g_2 )
  := \mathbf{m}_2 \; \exp\Big(\sum_{m,n,r \geq 1}
      t_{mnr}(\pa_{t_{mn}} \otimes \pa_{t_r}
      + \pa_{t_m} \otimes \pa_{t_{nr}})\Big) \, g_1 \otimes g_2 \, ,
\ee
with new deformation parameters $t_{mnr}$. This involves deformations  
not only with respect to the independent variables $t_m$, but also the previously introduced deformation parameters $t_{rs}$. 
It indeed defines an associative product.\footnote{The associativity can be checked 
as follows. First one verifies the identities
$P(\mathbf{m}_2 \otimes \id) = (\mathbf{m}_2 \otimes \id)( P_{13}+P_{23}
    + \sum t_{mnr}\pa_{t_m} \otimes \pa_{t_n}\otimes\pa_{t_r} )$
and $P(\id \otimes \mathbf{m}_2) = (\id \otimes \mathbf{m}_2)( P_{12}+P_{13}
    + \sum t_{mnr}\pa_{t_m} \otimes \pa_{t_n}\otimes\pa_{t_r} )$ where
$P:=\sum t_{mnr}(\pa_{t_{mn}}\otimes \pa_{t_r} + \pa_{t_m}\otimes \pa_{t_{nr}})$,
and, e.g., $P_{13}$ is given by $P$ acting on the first and third component of a
threefold tensor product (and as identity on the second). Writing
$\mathbf{m}_3 (\mathbf{m}_3 \otimes \id) = \mathbf{m}_2
e^P (\mathbf{m}_2 \otimes \id)(e^P \otimes \id)$, using the first identity, then
the associativity condition $\mathbf{m}_2 (\mathbf{m}_2 \otimes \id) = \mathbf{m}_2 (\id \otimes \mathbf{m}_2)$ for $\mathbf{m}_2$, and in the next step the second identity,
we obtain the associativity condition $\mathbf{m}_3 (\mathbf{m}_3 \otimes \id) = \mathbf{m}_3 (\id \otimes \mathbf{m}_3)$. }
There is a corresponding generalized Leibniz rule,
\be
    \pa_{t_{mnr}}(g_1\ast_3 g_2) = (\pa_{t_{mnr}}g_1)\ast_3 g_2 + (\pa_{t_{mn}} g_1)
    \ast_3 (\pa_{t_r} g_2) + (\pa_{t_m}g_1) \ast_3 (\pa_{t_{nr}} g_2)
    + g_1 \ast_3 (\pa_{t_{mnr}}g_2) \; .
\ee
Assuming $Z_{t_{mnr}}=0$ in addition to the linear system (\ref{Z_lin_sys}) and 
$Z_{t_{mn}}=0$, we are led to the deformation equations
\be
  \phi_{t_{mnr}} = - \phi_{t_{mn}} \ast_3 ( R_r + Q_r \phi ) \, ,
\ee
for which we find the following WNA version, 
\be
            f_{t_{mnr}} 
        &=& f_{t_{mn}} \ci_r f
         =  (f_{t_m} \ci_n f) \ci_r f 
         =  ((f \ci_m f) \ci_n f) \ci_r f  \nonumber \\
        &=& f \ci_m (f \ci_n (f \ci_r f)) - f \ci_{m+n} (f \ci_r f)
            - f \ci_m (f \ci_{n+r} f) + f \ci_{m+n+r} f \, ,
\ee
where now $A \ci_1 B = A \ast_3 QB$. 
Again, these flows commute with all other flows and among themselves conditionally, 
which now involves equations from (\ref{f_tn}) as well as from (\ref{f_tmn}).
\vskip.2cm

\noindent
\textbf{Remark.} If $g_1,g_2,g_3$ 
do not depend on the parameters $t_{kl}$ and $t_{mnr}$, then we have 
\be
  (g_1 \ast_3 g_2 \ast_3 g_3)_{t_{mnr}} = g_{1,t_m} \ast_3 g_{2,t_n} \ast_3 g_{3,t_r} \; .
\ee
As a consequence, 
\be
    \{g_1,g_2,g_3\}_{\ast_3} = \sum_{\sigma \in S_3} 
    \mathrm{sign}(\sigma) \, g_{\sigma(1)} \ast_3 g_{\sigma(2)} \ast_3 g_{\sigma(3)} 
\ee
yields an apparently new deformation quantization of the canonical Nambu bracket 
of order 3 (see  
\cite{Namb73,Takh94,DFST97,ALMY01,Curt+Zach03}, for example). 
See also \cite{Hiet97} for Nambu brackets in the context of integrable systems. 
\hfill $\square$
\vskip.2cm

The above procedure can be continued by setting\footnote{The remarks in 
footnote~\ref{foot:tensor-prod} apply correspondingly here. For an 
alternative but equivalent definition of these products see \cite{DMH05KPalgebra}. 
Exploiting equivalence transformations of star products (see \cite{Gutt+Rawn99}, 
for example), some parameters appear to be redundant. However, also deformation 
parameters which are redundant in this sense can play a non-trivial role e.g. 
as evolution variables of nonlinear integrable equations, see section~\ref{section:def->KP}. } 
\be
    \mathbf{m}_{n+1} = \mathbf{m}_n \exp\Big( \sum_{m_1,\ldots,m_{n+1} \geq 1}
     t_{m_1\ldots m_{n+1}} \sum_{r=1}^n \pa_{t_{m_1 \ldots m_r}}
      \otimes \pa_{t_{m_{r+1} \ldots m_{n+1}}} \Big)
     \qquad n=1,2,\ldots  \; .
\ee
It is not difficult to verify that these products are indeed associative (see also  \cite{DMH05KPalgebra}, appendix~C).

\begin{proposition}
The linear system (\ref{Z_lin_sys}) supplemented 
by $Z_{t_{m_1 \ldots m_k}} = 0$, $2 \leq k \leq n$, implies the 
Riccati system 
$\phi_{t_r} = S_r + L_r \phi - \phi R_r - \phi \ast_n Q_r \phi$, $r=1,2,\ldots$, 
supplemented by
\be
   \phi_{t_{m_1 \ldots m_k}} 
  = - \phi_{t_{m_1 \ldots m_{k-1}}} \ast_n (R_{m_k} + Q_{m_k} \phi) 
   \qquad \quad  k = 2,\ldots,n       \label{phi_t...}
\ee
(where $m_1,\ldots,m_k = 1,2,\ldots$). 
\end{proposition}
\textbf{Proof:} This is a straightforward generalization of the proof 
of proposition~\ref{prop:Ztmn-equiv}. 
\hfill $\square$
\vskip.2cm

The extended Riccati system in the last proposition translates into
\be
     f_{t_{m_1 \ldots m_k}} = e_{(m_1, \ldots, m_k)}  
        \qquad \quad     k = 1,\ldots,n  \, , 
\ee
where
\be
    e_{(m_1, \ldots, m_k)} := R_f^{(m_k)} \cdots R_f^{(m_1)} (f)
\ee
and $R_f^{(m)}(a) := a \ci_m f$, hence  
$e_{(m_1,\ldots,m_k)} \ci_{m_{k+1}} f = e_{(m_1,\ldots,m_{k+1})}$. 
The products $\ci_m$ involve the $\ast_n$ product. It is convenient to replace 
it by the $n \to \infty$ limit of the $\ast_n$ products and this will be done 
in the following.

\begin{lemma}
Let $f_{t_{m_1 \cdots m_k}} \in \A'$ for $k=1,\ldots,r$.  
The partial derivative with respect to $t_{m_1 \ldots m_r}$ then 
satisfies the generalized Leibniz rule 
\bez
     (a \ci_n b)_{t_{m_1 \cdots m_r}} 
 &=& \sum_{k=0}^r a_{t_{m_1 \cdots m_k}} \ci_n b_{t_{m_{k+1} \cdots m_r}} \nonumber \\
 &=& a \ci_n b_{t_{m_1 \cdots m_r}}
     + \sum_{k=1}^{r-1} a_{t_{m_1 \cdots m_k}} \ci_n b_{t_{m_{k+1} \cdots m_r}} 
     + a_{t_{m_1 \cdots m_r}} \ci_n b
\eez
with respect to all products $\ci_n$, $n=1,2,\ldots$, defined by (\ref{circ_n}) 
in terms of $f$. 
\end{lemma}
{\em Proof:} This holds for $n=1$ (cf. also \cite{DMH05KPalgebra}, appendix~C). 
The general statement is then obtained by induction as follows. 
By use of (\ref{circ_n}) and the induction hypothesis, we find
\bez
   (a \ci_{n+1} b)_{t_{m_1 \cdots m_r}} 
 &=& \sum_{0 \leq k \leq l \leq r} \Big(
   a_{t_{m_1 \cdots m_k}}\ci_1 ( f_{t_{m_{k+1} \cdots m_l}} \ci_n b_{t_{m_{l+1} 
       \cdots m_r}} ) \\
 & & - (a_{t_{m_1 \cdots m_k}} \ci_1  f_{t_{m_{k+1} \cdots m_l}}) 
         \ci_n b_{t_{m_{l+1} \cdots m_r}} \Big) \; .
\eez
Because of our assumption and the WNA property of the products (see proposition~3.1 
in \cite{DMH06nahier}), only the terms with $k=l$ survive. Since 
$f_{t_{m_{k+1} \cdots m_k}}$ has to be read as $f$, this leads to
\bez
    (a \ci_{n+1} b)_{t_{m_1 \cdots m_r}} 
 &=& \sum_{0 \leq k \leq r} \Big(
   a_{t_{m_1 \cdots m_k}} \ci_1 ( f  \ci_n b_{t_{m_{k+1} \cdots m_r}} ) 
   - (a_{t_{m_1 \cdots m_k}} \ci_1 f) \ci_n b_{t_{m_{k+1} \cdots m_r}} \Big) \\
 &=& \sum_{0 \leq k \leq r} a_{t_{m_1 \cdots m_k}} \ci_{n+1} b_{t_{m_{k+1} \cdots m_r}}  
     \; .
\eez
\hfill $\square$

\begin{proposition}
\label{prop:genhier_comm}
The flows given by
\be
    f_{t_{m_1 \ldots m_n}} = e_{(m_1, \ldots, m_n)}  
      \qquad \quad n=1,2,\ldots, \quad m_k \in \mathbb{N}   \label{genhier}
\ee
commute with each other.
\end{proposition}
{\em Proof:} We use induction. For $n=1$ the statement is contained 
in theorem~\ref{theorem:WNA}. 
Using $e_{(m_1,\ldots, m_{r+1})} = e_{(m_1,\ldots,m_r)} \ci_{m_{r+1}} f$ and 
the preceding lemma (the assumptions of which are satisfied as a consequence 
of (\ref{genhier})), we obtain
\bez
 & &   f_{t_{m_1 \cdots m_{r+1}} t_{n_1 \cdots n_{s+1}}} 
  = e_{(m_1,\ldots, m_{r+1}),t_{n_1 \cdots n_{s+1}}} \\
 &=& e_{(m_1,\ldots, m_r),t_{n_1 \cdots n_{s+1}}}\ci_{m_{r+1}} f
     + \sum_{k=0}^s  e_{(m_1,\ldots, m_r),t_{n_1 \cdots n_k}} \ci_{m_{r+1}} 
          e_{(n_{k+1},\ldots, n_{s+1})} \, ,
\eez
where we made use of some `lower' equations of the system (\ref{genhier}). 
By induction hypothesis, the first summand of the last expression is equal to  
\bez
 & &   e_{(n_1,\ldots, n_{s+1}),t_{m_1 \cdots m_r}} \ci_{m_{r+1}} f  \\
 &=& \Big( e_{(n_1,\ldots, n_s),t_{m_1 \cdots m_r}} \ci_{n_{s+1}} f
     + \sum_{k=0}^{r-1} e_{(n_1,\ldots, n_s),t_{m_1 \cdots m_k}} 
        \ci_{n_{s+1}} e_{(m_{k+1},\ldots, m_r)} \Big) \ci_{m_{r+1}} f \\
 &=& (e_{(n_1,\ldots, n_s),t_{m_1 \cdots m_r}} \ci_{n_{s+1}} f) \ci_{m_{r+1}} f
     + \sum_{k=0}^{r-1} e_{(n_1,\ldots, n_s),t_{m_1 \cdots m_k}} 
         \ci_{n_{s+1}} e_{(m_{k+1},\ldots, m_{r+1})} \\
 &=& (e_{(n_1,\ldots, n_s),t_{m_1 \cdots m_r}} \ci_{n_{s+1}} f) \ci_{m_{r+1}} f 
     - e_{(n_1,\ldots, n_s),t_{m_1 \cdots m_r}} \ci_{n_{s+1}} (f \ci_{m_{r+1}} f) \\
 & & + \sum_{k=0}^r e_{(n_1,\ldots, n_s),t_{m_1 \cdots m_k}} 
         \ci_{n_{s+1}} e_{(m_{k+1},\ldots, m_{r+1})}  \\
 &=& - e_{(n_1,\ldots, n_s),t_{m_1 \cdots m_r}} \ci_{(m_{r+1}+n_{s+1})} f 
     + \sum_{k=0}^r e_{(n_1,\ldots, n_s),t_{m_1 \cdots m_k}} 
         \ci_{n_{s+1}} e_{(m_{k+1},\ldots, m_{r+1})} \, , 
\eez
by use of proposition~3.3 in \cite{DMH06nahier}. 
Inserting this in the above expression for 
$f_{t_{m_1 \cdots m_{r+1}} t_{n_1 \cdots n_{s+1}}}$, and applying the 
induction hypothesis to $e_{(n_1,\ldots, n_s),t_{m_1 \cdots m_r}}$, 
results in an expression which is symmetric in $m$'s and $n$'s. 
This implies our assertion. 
\hfill $\square$
\vskip.2cm

For $n=1$, (\ref{genhier}) is the hierarchy (\ref{f_tn}). 
Whereas any two flows of the latter system commute with each other 
without further conditions, the commutativity of two flows of the 
extended system (\ref{genhier}) on the $n$th level, $n>1$, also requires 
some flow equations on lower levels, as explained above. 
The structure of the proof of the last proposition clearly displays this fact. 
Because of this conditional commutativity, (\ref{genhier}) is \emph{not} 
quite a hierarchy in the familiar sense, though a natural generalization, 
which obviously preserves the integrability of the hierarchy. 
\vskip.2cm

By elimination of $L,R,S$ from (\ref{phi_t...}) (using also the 
Riccati system (\ref{phi_tn})), respectively elimination of bare $f$'s 
from (\ref{genhier}), one obtains deformation equations of the 
deformed (matrix) pKP hierarchy by an extension of the methods 
in section~\ref{section:WNA-Moyal} (see also \cite{DMH05KPalgebra} 
for a different approach). Again, these deformation equations extend the 
deformed pKP hierarchy. 
\vskip.2cm

\noindent
\textbf{Remark.} 
We know from \cite{DMH06nahier} that the monomial $e_{(m_1,\ldots,m_n)}$
is related to a basis element of the space of quasi-symmetric functions 
(see \cite{Gess84,Reut93,Haze00}, for example) 
in commuting variables $k_1,k_2,\ldots,k_N$, given by
\be
    M_{(m_1,\ldots,m_n)} = \sum_{1 \leq r_1 < r_2 < \cdots < r_n \leq N}
        k_{r_1}^{m_1} k_{r_2}^{m_2} \cdots k_{r_n}^{m_n} \; .
\ee
The relation of the more general deformations, considered in this section, 
with quasi-symmetric functions emerges as follows on a simpler level. 
Products of the form
\be
    e^{\xi(k_1)} \ast_n e^{\xi(k_2)} \ast_n \cdots \ast_n e^{\xi(k_N)}
  =: e^{\Xi(k_1,\ldots,k_N)}    \label{e^xi_ast_prod}
\ee
with $\xi(k) = \sum_{n \geq 1} t_n k^n$ are constituents of deformed
soliton solutions as obtained in \cite{DMH05KPalgebra}. We find\footnote{Note 
that $M_{(m_1,\ldots, m_r)}=0$ if $r>N$. 
See also section~8.1 of \cite{DMH05KPalgebra}.}
\be
    \Xi(k_1,\ldots,k_N)
  = \sum_{r=1}^{n+1} \, \sum_{m_1,\ldots,m_r \geq 1}
     t_{m_1 \ldots m_r} \, M_{(m_1,\ldots,m_r)}  \; .
\ee
\hfill $\square$

\section{Via deformation to KP}
\label{section:def->KP}
\setcounter{equation}{0}
Let us start with an associative algebra where the elements 
depend smoothly on (only) $t_1$. The deformations considered in 
section~\ref{section:beyond} then reduce to those associated with the 
deformation parameters  
\be
     t_{1^k} := t_{1 \ldots 1}  \qquad \quad k=2,3,\ldots \; .
\ee
Let $\ast$ denote the $n \to \infty$ limit of the corresponding products $\ast_n$. 
The set of derivatives $\{ \pa_{t_{1^k}} \}_{k \in \mathbb{N}}$ is a 
Hasse-Schmidt derivation of the product $\ast$, i.e. they satisfy the 
generalized derivation rule
\be
   (g_1 \ast g_2)_{t_{1^k}} 
  = \sum_{j=0}^{k} g_{1,t_{1^j}} \ast g_{2,t_{1^{k-j}}}
  = g_1 \ast g_{2,t_{1^k}} 
    + \sum_{j=1}^{k-1} g_{1,t_{1^j}} \ast g_{2,t_{1^{k-j}}} 
    + g_{1,t_{1^k}} \ast g_2 \; .    \label{HS-deriv}
\ee
Given commuting (ordinary) derivations $\eth_k$, $k=1,2,\ldots$, the set 
$\{ (-1)^k \mathbf{p}_k(-\tilde{\eth}) \}_{k \in \mathbb{N}}$, with 
the elementary Schur polynomials $\mathbf{p}_k$ and
$\tilde{\eth} = (\eth_{t_1}, \frac{1}{2} \eth_{t_2}, \frac{1}{3} \eth_{t_3}, \ldots)$, 
also yields a Hasse-Schmidt derivation. This observation allows us to 
\emph{define} derivations $\eth_k$, $k=1,2,\ldots$, by setting
\be
    \pa_{t_{1^k}} =: (-1)^k \mathbf{p}_k(-\tilde{\eth})  \qquad k=1,2,\ldots \; .
               \label{pa_t1^k-id}
\ee
In particular, we have $\eth_1 = \pa_{t_1}$ and $\eth_2 = \pa_{t_1}^2 - 2 \pa_{t_{11}}$. 
\vskip.2cm

Now, what has all this to do with KP?
According to our general strategy, we should start from the linear equation 
\be
       Z_{t_1} = H \, Z    \label{lin_sys_t1}
\ee
and assume $Z$ to be independent of the deformation parameters, i.e.,\footnote{By 
use of (\ref{pa_t1^k-id}), the system (\ref{pre_heat_hier}) is equivalent to 
$\eth_n(Z) = \pa_{t_1}^n(Z)$, $n=2,3,\ldots$. Via $\eth_n \mapsto \pa_{t_n}$, 
which is the effect of the homomorphism $\hat{\,}$ introduced below, 
this becomes the linear heat hierarchy $Z_{t_n} = \pa_{t_1}^n(Z)$. }
\be
     Z_{t_{1^n}} = 0 \qquad \quad n=2,3,\ldots \; .  \label{pre_heat_hier}
\ee
By use of the last condition, (\ref{HS-deriv}), and $X_{t_1} = R X + Q Y$ (which 
follows from (\ref{lin_sys_t1}) with the decomposition (\ref{Z,H-decomp})), 
we obtain
\be
    \phi_{t_{1^{n+1}}} 
   = - \sum_{k=0}^n \phi_{t_{1^k}} \ast X_{t_{1^{n-k}}} \ast X^{\ast-1} 
   = - \phi_{t_{1^n}} \ast X_{t_1} \ast X^{\ast-1}
   = - \phi_{t_{1^n}} \ast (Q \phi + R) \; .
\ee
This in turn leads to 
\be
    \phi_{t_{1^{n+1}}t_{1^m}} = - \phi_{t_{1^n}t_{1^m}} \ast(Q \phi +R) 
    - \sum_{k=0}^{m-1} \phi_{t_{1^n} t_{1^k}} \ast Q \phi_{t_{1^{m-k}}} \, ,
\ee
and consequently, by elimination of $R$, 
\be
    \phi_{t_{1^{n+1}}t_{1^m}} - \phi_{t_{1^{m+1}}t_{1^n}}
  = \sum_{k=0}^{n-1} \phi_{t_{1^m} t_{1^k}} \ast Q \phi_{t_{1^{n-k}}}
    - \sum_{k=0}^{m-1} \phi_{t_{1^n} t_{1^k}} \ast Q \phi_{t_{1^{m-k}}} \, , 
          \label{pKP_in_disguise}
\ee
which by use of (\ref{pa_t1^k-id}) results in 
\be
    && \mathbf{p}_{n+1}(-\tilde{\eth}) \mathbf{p}_m(-\tilde{\eth}) \phi 
  - \mathbf{p}_{m+1}(-\tilde{\eth}) \mathbf{p}_n(-\tilde{\eth}) \phi \nonumber \\
    &=& \sum_{k=0}^{m-1} \mathbf{p}_k(-\tilde{\eth}) \mathbf{p}_n(-\tilde{\eth}) 
        \phi \ast Q \mathbf{p}_{m-k}(-\tilde{\eth}) \phi
  - \sum_{k=0}^{n-1} \mathbf{p}_k(-\tilde{\eth}) \mathbf{p}_m(-\tilde{\eth}) 
        \phi \ast Q \mathbf{p}_{n-k}(-\tilde{\eth})\phi \, , 
              \label{pKP_in_disguise2}
\ee
where $m,n=1,2,\ldots$. 
Abstracting from the origin of $\phi$ and regarding this as a system 
imposed on some $\phi(t_1,t_{11},\ldots)$, it resembles a known formulation 
of the pKP hierarchy \cite{DNS89,DMH06nahier}, though with the deformed product 
and with partial derivatives with respect to variables $t_2,t_3,\ldots$ replaced 
by $\eth_k$, $k=2,3,\ldots$. 
Since the latter are derivations of the product $\ast$, the operator defined by 
\be
   \hat{g} := \exp( \sum_{k=2}^n t_k \eth_k ) \, g 
\ee
is a homomorphism, which satisfies
\be
   \hat{g}_{t_k} = \widehat{\eth_k(g)} \; .
\ee
Applying this operator to (\ref{pKP_in_disguise2}), and 
finally setting all deformation parameters to zero, we obtain for  
\be
  \hat{\phi} := \exp\Big( \sum_{k=2}^n t_k \eth_k \Big) \, \phi \Big|_{t_{1^k}=0, \, k>1} 
        \label{hatphi}
\ee
the pKP hierarchy equations
\be
    && \mathbf{p}_{n+1}(-\tilde{\pa}) \mathbf{p}_m(-\tilde{\pa}) \hat{\phi} 
  - \mathbf{p}_{m+1}(-\tilde{\pa}) \mathbf{p}_n(-\tilde{\pa}) \hat{\phi} \nonumber \\
    &=& \sum_{k=0}^{m-1} \mathbf{p}_k(-\tilde{\pa}) \mathbf{p}_n(-\tilde{\pa}) 
        \hat{\phi} Q \mathbf{p}_{m-k}(-\tilde{\pa}) \hat{\phi}
  - \sum_{k=0}^{n-1} \mathbf{p}_k(-\tilde{\pa}) \mathbf{p}_m(-\tilde{\pa}) 
        \hat{\phi} Q \mathbf{p}_{n-k}(-\tilde{\pa})\hat{\phi} \; . 
\ee
Any solution $\phi(t_1,t_{11},\ldots)$ of (\ref{pKP_in_disguise}) thus yields, 
via (\ref{hatphi}), a solution of the pKP hierarchy (in the matrix algebra 
with product modified by $Q$). 
\vskip.1cm

The way towards solutions of the pKP hierarchy described above is surprisingly simple. 
We have to write down the solution of the single linear ordinary differential 
equation (\ref{lin_sys_t1}), which is a matrix of functions of the variable $t_1$.
All the other variables, the evolution parameters of the hierarchy, emerge 
in the pKP solution $\phi$ as deformation parameters through the product!  
\vskip.2cm

\noindent
\textbf{Example.} Let $S=0$ and $Q = RK-KL$ with an $N \times M$ matrix $K$ 
(cf. (\ref{Q=RK-KL})). 
Then the solution of the linear equation (\ref{lin_sys_t1}) takes the form
\be
    Z = e^{t_1 H} Z_0 
      = \left( \begin{array}{c} 
          e^{t_1 R} X_0 + (e^{t_1 R} K - K e^{t_1 L}) Y_0 \\
          e^{t_1 L} Y_0
               \end{array} \right) \, ,
\ee
which leads to the solution\footnote{Note that the $\ast$-inverse of a function 
of $t_1$ depends on all deformation parameters. For example, the inverse of 
$e^{t_1 L}$ is given by $e^{-t_1 L + t_{11} L^2 - t_{111} L^3 + \ldots}$.} 
\be
   \phi = e^{t_1 L} \phi_0 \ast \Big( e^{t_1 R} (I_N + K \phi_0)  
          - K e^{t_1 L} \phi_0 \Big)^{\ast -1}  \, , 
\ee
of (\ref{pKP_in_disguise}), where $\phi_0 := Y_0 X_0^{-1}$. This 
expression is solely built from functions of $t_1$ only. The additional variables 
enter via the deformed product. 
Using the homomorphism property of the operation $\hat{\,}$, we recover the 
solution (\ref{phi_solution}) of the pKP hierarchy. 
\hfill $\square$
\vskip.2cm

\noindent
\textbf{Remark.} The deformation considered in this section is trivial in the sense 
that it can be transformed away by an equivalence transformation of star products. 
This will be demonstrated in the following. Consider the product
\be
    g_1 \star_n g_2 
 := \mathcal{D}_n^{-1} ( \mathcal{D}_n(g_1) \ast_{n-1} \mathcal{D}_n(g_2) ) \, , 
\ee
with
\be
    \mathcal{D}_n = e^{- t_{1^n} \, \Upsilon_n } \, ,
\ee
where $\Upsilon_n$ is an operator that commutes with $t_{1^n}$. Then
\be
     (g_1 \star_n g_2)_{t_{1^n}}
  = \Upsilon_n ( g_1 \star_n g_2 ) - \Upsilon_n(g_1) \star_n g_2 
     - g_1 \star_n \Upsilon_n(g_2) + g_{1,t_{1^n}} \star_n g_2
     + g_1 \star_n g_{2,t_{1^n}}  \; .
\ee
Now let us choose
\be
    \Upsilon_n = (-1)^n \mathbf{p}_n(-\tilde{\eth}) + \frac{1}{n} (-1)^n \eth_n \; .
\ee
Here the last term cancels a corresponding term in 
$(-1)^n \mathbf{p}_n(-\tilde{\eth}) = \pa_{t_{1^n}}$, so that the remaining part  
can be expressed in terms of partial derivatives with respect to $\pa_{t_{1^k}}$ 
with $k<n$ only. Using the derivation property of $\eth_n$, we find
\be
     (g_1 \star_n g_2)_{t_{1^n}}
 &=& (-1)^n \mathbf{p}_n(-\tilde{\eth}) ( g_1 \star_n g_2 ) 
     - (-1)^n \mathbf{p}_n(-\tilde{\eth})(g_1) \star_n g_2 
     - g_1 \star_n (-1)^n \mathbf{p}_n(-\tilde{\eth})(g_2) \nonumber \\ 
 & & + g_{1,t_{1^n}} \star_n g_2 + g_1 \star_n g_{2,t_{1^n}}  \nonumber \\
 &=& \sum_{k=0}^n \pa_{t_{1^k}}(g_1) \star_n \pa_{t_{1^{n-k}}}(g_2) 
\ee
with the help of the (Hasse-Schmidt) generalized derivation property of 
$(-1)^n \mathbf{p}_n(-\tilde{\eth})$, and (\ref{pa_t1^k-id}). 
This is precisely the expression (\ref{HS-deriv}) with products $\star_n$ 
and $\ast_n$ exchanged, and this also holds for the iterated derivatives. 
Since these expressions coincide for both products at $t_{1^n} =0$, we conclude 
that the two products are the same. We have shown that $\ast_n$ is obtained 
from $\ast_{n-1}$ by an equivalence transformation of star products, 
in the case where only the deformation parameters $t_{1^n}$, $n>1$, are switched on.
\hfill $\square$
\vskip.2cm

\noindent
\textbf{Remark.} With the (restricted) deformation considered in this section, we have 
\be
    e^{t_1 k_1} \ast \cdots \ast e^{t_1 k_N} = e^{\Xi(k_1,\ldots,k_N)}
\ee
with
\be
   \Xi(k_1,\ldots,k_N) = t_1 M_{(1)} + t_{11} M_{(1,1)} + \ldots + t_{1^N} M_{(1^N)} 
\ee
(see the last remark in section~\ref{section:beyond}). The expressions $M_{(1^k)}$, $k=1,\ldots,N$, are the elementary symmetric functions in $k_1,\ldots,k_N$. 
These are related to the power symmetric functions $M_{(k)}$ through the relations
\be
    M_{(1^n)} = (-1)^n \mathbf{p}_n(-\tilde{M}) \qquad n=1,2,\ldots 
   \, , \qquad
    \tilde{M} := (M_{(1)}, M_{(2)}/2, M_{(3)}/3,\ldots) 
\ee
(cf. \cite{Macd95}, the second of equations (2.14') in chapter~I). 
Thus, via the product $\ast$, elementary symmetric functions $M_{(1^k)}$ are 
introduced in our expressions. After use of the last relation to express the 
elementary symmetric functions in terms of power symmetric functions, the 
hat operator then generates the new coefficients $t_2,t_3,\ldots$ via 
\be
    \hat{\Xi}(k_1,\ldots,k_N) 
  = \Xi(k_1,\ldots,k_N) + t_2 M_{(2)} + \cdots + t_N M_{(N)} \; .
\ee
Elementary symmetric functions no longer show up in the final 
expressions, after setting $t_{1^k}=0$, $k>1$. 
\hfill $\square$

\section{Conclusions}
\label{section:conclusions}
\setcounter{equation}{0}
In the presence of a Moyal or (in the sense of section~\ref{section:beyond}) 
generalized deformation, the (potential) KP hierarchy extends to a larger hierarchy 
which includes additional flows with the deformation parameters as new evolution 
variables. This observation has already been made in previous work 
(see \cite{DMH05KPalgebra}, in particular). Here we presented a new and quite 
simple approach which also yields exact solutions. 
\vskip.1cm

As already pointed out in section~\ref{section:Moyal}, in this case we are 
actually dealing with a \emph{family} of KP hierarchies, parametrized 
by deformation parameters of the product. From the analogy with 
Seiberg-Witten maps it should be obvious that a deformation equation does 
\emph{not} generate a homomorphism of an algebra, but rather a homomorphism 
between different algebras (corresponding to different non-zero values of the 
deformation parameters). This is mirrored in the fact that the derivatives 
$\pa_{t_{m_1\ldots m_r}}$, $r>1$, are \emph{not} derivations. 
\vskip.1cm

In \cite{DMH06nahier} we proved that any solution of the hierarchy (\ref{f_tn}) of 
ordinary differential equations in any WNA algebra $\A$ determines a solution of 
the potential KP hierarchy in the middle nucleus $\A'$ of $\A$. 
The present work extended the hierarchy (\ref{f_tn}) by corresponding ordinary 
differential equations in the deformation parameters. 
\vskip.1cm

Concerning the significance of an extended hierarchy, we already made some remarks 
in the introduction, but this work further adds to it. 
In section~\ref{section:def->KP} we considered a sub-hierarchy 
of deformation equations which turned out to be the KP hierarchy in disguise. 
In this example, the KP hierarchy equations themselves emerge as deformation flow equations. Although the deformation is trivial in the sense that the deformation 
of the product can be eliminated by an equivalence transformation, the 
deformation flows act in a non-trivial way. 
\vskip.1cm

There is certainly much more to be discovered in the huge extended hierarchy 
with the general deformation treated in section~\ref{section:beyond}. The fact 
that Moyal products exhibit structures of relevance in the theory of integrable 
systems (see \cite{Olve+Sand00,Atho07}, for example) provides us with further 
motivation. In fact, the results (of section~\ref{section:def->KP}) just 
mentioned demonstrate, in a completely independent way, the relevance of more 
general star products in this respect. 
\vskip.1cm

The FORM programs listed in the appendices should also be helpful to work 
out further examples beyond those presented in this work.

\renewcommand{\theequation} {\Alph{section}.\arabic{equation}}

\section*{Appendix A: Checking commutativity of flows with FORM in the KdV case }
\setcounter{section}{1}
\setcounter{equation}{0}
In this appendix we list a simple FORM program \cite{Heck00,Verm02} which 
verifies that the flow of the Moyal-deformed (matrix) KdV equation (\ref{ncKdV}) 
commutes with that of its deformation equation (\ref{ncKdV_def}). 
The program can easily be adapted to other equations. 
FORM is especially helpful for such computations since by default it deals 
with \emph{noncommuting} functions. In the following, an expression like 
$\tt u(3,2,1)$ stands for $u_{xxxtt\theta}$. 

\begin{verbatim}
* FORM program ( see http://www.nikhef.nl/~form/ )
* Define (noncommuting) objects:
Function u,dx;
Symbol x,t,th;
* right hand side of (deformed) KdV equation 
Local ut = - u(3,0,0) + 3*( u(1,0,0)*u(0,0,0) + u(0,0,0)*u(1,0,0) );
* and derivatives with respect to x:
Local utx = dx*ut;
Local utxx = dx*utx;
* rhs of deformation equation and derivatives w.r. to x:
Local uth = ( u(2,0,0)*u(0,0,0) - u(0,0,0)*u(2,0,0) )/2;
Local uthx = dx*uth;
Local uthxx = dx*uthx;
Local uthxxx = dx*uthxx;
repeat;
* partial differentiation w.r. to x:
  id dx*u(x?,t?,th?) = u(x+1,t,th) + u(x,t,th)*dx;
endrepeat;
  id dx = 0;
.sort

Function dt,dth;
* Check of commutativity of flows. The following should vanish.
Local diff = dt*uth - dth*ut;
repeat;
* differentiation w.r. to the deformation parameter:
  id dth*u(x?,t?,th?) =  u(x,t,th+1) + u(x,t,th)*dth
       + (1/2)*( u(x+1,t,th)*dt - u(x,t+1,th)*dx );
* partial differentiation w.r. to x and t: 
  id dx*u(x?,t?,th?) = u(x+1,t,th) + u(x,t,th)*dx;
  id dt*u(x?,t?,th?) = u(x,t+1,th) + u(x,t,th)*dt;
* rhs of ncKdV equation and derivatives w.r. to x:
  id u(0,1,0) = ut; id u(1,1,0) = utx; id u(2,1,0) = utxx;
* rhs of deformation equation and derivatives w.r. to x:
  id u(0,0,1) = uth; id u(1,0,1) = uthx; id u(3,0,1) = uthxxx;
endrepeat;
  id dx?{dx,dt,dth} = 0;
print diff;
.end
\end{verbatim}

\section*{Appendix B: Checking a deformation equation with FORM}
\setcounter{section}{2}
\setcounter{equation}{0}
The following FORM program verifies that the 
right hand side of the deformation equation (\ref{phi_thmn}) with $m=1$ and $n=2$ 
becomes the right hand side of (\ref{phi_th12}) by elimination of $L,R,S$, 
using the first three equations of the (Moyal-deformed) Riccati system (\ref{phi_tn}). 
Here we set $x=t_1, y=t_2, t=t_3$.

\begin{verbatim}
Function phi,phix,phiy,phit,L,Q,R,S,dx;
Symbol n;
Local rhs1 = (phit - dx*dx*phix)/6 - phix*Q*phix;
Local rhs2 = ( phiy*(R+Q*phi) - phix*(R(2)+Q(2)*phi) )/2;
Local diff = rhs1 - rhs2;
repeat;
* the Riccati equations:
  id phix = S + L*phi - phi*R - phi*Q*phi;
  id phiy = S(2) + L(2)*phi - phi*R(2) - phi*Q(2)*phi;
  id phit = S(3) + L(3)*phi - phi*R(3) - phi*Q(3)*phi;
* differentiation rule:
  id dx*phi = phix + phi*dx;
* L,Q,R,S are constant:
  id dx*L?{L,Q,R,S} = L*dx;
* recursion relations:
  id L(n?{2,3}) = L*L(n-1)+S*Q(n-1);
  id Q(n?{2,3}) = Q*L(n-1)+R*Q(n-1);
  id R(n?{2,3}) = Q*S(n-1)+R*R(n-1);
  id S(n?{2,3}) = L*S(n-1)+S*R(n-1);
  id L?{L,Q,R,S}(1) = L;
endrepeat;
.sort
id dx = 0;
print diff;
.end
\end{verbatim}


\begin{thebibliography}{10}

\bibitem{BFFLS78a}
Bayen F, Flato M, Fronsdal C, Lichnerowicz A, and Sternheimer D 1978 
Deformation theory and quantization. I. Deformation of symplectic
  structures 
{\em Ann. Phys.} {\bf 111} 61--110

\bibitem{Doug+Nekr01}
Douglas M R and Nekrasov N A 2001 
Noncommutative field theory 
{\em Rev. Mod. Phys.} {\bf 73} 977--1029

\bibitem{Kupe90}
Kupershmidt B A 1990 
Quantizations and integrable systems 
{\em Lett. Math. Phys.} {\bf 20} 19--31

\bibitem{FMR93}
Figueroa-O'Farrill J M, Mas J and Ramos E 1993 
A one-parameter family of Hamiltonian structures for the KP
  hierarchy and a continuous deformation of the nonlinear $W_{KP}$ algebra 
{\em Commun. Math. Phys.} {\bf 158} 17--43

\bibitem{Stra95}
Strachan I 1995
The Moyal bracket and the dispersionless limit of the KP hierarchy
{\em J. Phys. A: Math. Gen.} {\bf 28} 1967--1976

\bibitem{Gawr95}
Gawrylczyk J 1995
Relationship between the Moyal KP and the Sato KP hierarchies
{\em J. Phys. A: Math. Gen.} {\bf 28} 4381--4388

\bibitem{Stra97}
Strachan I A B 1997 
A geometry for multidimensional integrable systems
{\em J. Geom. Phys.} {\bf 21} 255--278

\bibitem{Koik01}
Koikawa T 2001 
Soliton equations extracted from the noncommutative zero-curvature equation
{\em Progr. Theor. Phys.} {\bf 105} 1045--1057

\bibitem{Blas+Szab03}
B{\l}aszak M and Szablikowski B M 2003 
From dispersionless to soliton systems via Weyl-Moyal-like deformations 
{\em J. Phys. A: Math. Gen.} {\bf 36} 12181--12203 

\bibitem{Carr03qm}
Carroll R 2003
Integrable systems as quantum mechanics 
{\em Preprint} quant-ph/0309159

\bibitem{Taka94}
Takasaki K 1994
Nonabelian KP hierarchy with Moyal algebraic coefficients 
{\em J. Geom. Phys.} {\bf 14} 332--364

\bibitem{Taka01}
Takasaki K 2001
Anti-self-dual Yang-Mills equations on noncommutative spacetime 
{\em J. Geom. Phys.} {\bf 37} 291--306

\bibitem{Wang+Wada03ncsoliton}
Wang N and Wadati M 2003 
A new approach to noncommutative soliton equations 
{\em J. Phys. Soc. Japan} {\bf 72} 3055--3062 

\bibitem{Lech+Popo04}
Lechtenfeld O and Popov A D 2004 
Noncommutative monopoles and Riemann-Hilbert problems 
{\em JHEP} {\bf 01} 069 

\bibitem{Hama05}
Hamanaka M 2005
Noncommutative solitons and integrable systems 
{\em Preprint} hep-th/0504001 

\bibitem{DMH00ncKdV}
Dimakis A and M\"uller-Hoissen F 2000 
The Korteweg-de-Vries equation on a noncommutative space-time 
{\em Phys. Lett. A} {\bf 278} 139--145

\bibitem{DMH00SW}
Dimakis A and M\"uller-Hoissen F 2000
Moyal deformation, Seiberg-Witten map, and integrable models 
{\em Lett. Math. Phys.} {\bf 54} 123--135 

\bibitem{DMH01ncNLS}
Dimakis A and M\"uller-Hoissen F 2001 
Noncommutative NLS equation 
{\em Czech. J. Phys.} {\bf 51} 1285--1290

\bibitem{Wang04}
Wang N 2004
Rescaling symmetry flow of the Kadomtsev-Petviashvili hierarchy 
{\em Chin. Phys. Lett.} {\bf 21} 2327--2329 

\bibitem{DMH04hier}
Dimakis A and M\"uller-Hoissen F 2004 
Extension of noncommutative soliton hierarchies 
{\em J. Phys. A: Math. Gen.} {\bf 37} 4069--4084 

\bibitem{DMH04ncKP}
Dimakis A and M\"uller-Hoissen F 2004 
Explorations of the extended ncKP hierarchy 
{\em J. Phys. A: Math. Gen.} {\bf 37} 10899--10930 

\bibitem{DMH04extMoyal}
Dimakis A and M\"uller-Hoissen F 2004 
Extension of Moyal-deformed hierarchies of soliton equations 
{\em XI International Conference Symmetry Methods in Physics} 
ed \v{C} Burdik, O Navr{\'a}til and S Po\v{s}ta 
(Dubna: Joint Institute for Nuclear Research) 

\bibitem{DMH05KPalgebra}
Dimakis A and M\"uller-Hoissen F 2005 
An algebraic scheme associated with the noncommutative KP hierarchy 
and some of its extensions 
{\em J. Phys. A: Math. Gen.} {\bf 38} 5453--5505 

\bibitem{Seib+Witt99}
Seiberg N and Witten E 1999 
String theory and noncommutative geometry 
{\em JHEP} {\bf 9909} 032 

\bibitem{Heck00}
Heck A 2000
{\em FORM for Pedestrians} 
(Amsterdam: NIKHEF)

\bibitem{Verm02}
Vermaseren J A M 2002 
{\em FORM Reference Manual} 
(Amsterdam: NIKHEF) 

\bibitem{Sato+Sato82}
Sato M and Sato Y 1982 
Soliton equations as dynamical systems on infinite dimensional 
Grassmann manifold 
{\em Nonlinear Partial Differential Equations in Applied Science} 
ed H Fujita, P D Lax and G Strang
(Amsterdam: North-Holland) pp 259--271

\bibitem{Taka89}
Takasaki K 1989 
Geometry of universal Grassmann manifold from algebraic point of view 
{\em Rev. Math. Phys.} {\bf 1} 1--46 

\bibitem{DMH06nahier}
Dimakis A and M\"uller-Hoissen F 2006 
Nonassociativity and integrable hierarchies 
{\em Preprint} nlin.SI/0601001 

\bibitem{DMH07Ricc}
Dimakis A and M\"uller-Hoissen F 2007 
Weakly nonassociative algebras, Riccati and KP hierarchies 
{\em Preprint} nlin.SI/0701010 

\bibitem{DMH07Wronski}
Dimakis A and M\"uller-Hoissen F 2007 
With a Cole-Hopf transformation to solutions of the 
noncommutative KP hierarchy in terms of Wronski matrices 
{\em to appear in J. Phys. A: Math. Theor.} 

\bibitem{DMH06Burgers}
Dimakis A and M\"uller-Hoissen F 2006 
Burgers and KP hierarchies: A functional representation approach 
{\em to appear in Theor. Math. Phys.} [nlin.SI/0601001] 

\bibitem{FMP98}
Falqui G, Magri F and Pedroni M 1998 
Bihamiltonian geometry, Darboux coverings, and linearization of the 
KP hierarchy 
{\em Commun. Math. Phys.} {\bf 197} 303--324 

\bibitem{GMA94}
Guil F, Ma{\~n}as M and {\'A}lvarez G 1994 
The Hopf-Cole transformation and the KP equation 
{\em Phys. Lett. A} {\bf 190} 49--52 

\bibitem{DMH06CJP}
Dimakis A and M\"uller-Hoissen F 2006 
From nonassociativity to solutions of the KP hierarchy 
{\em Czech. J. Phys.} {\bf 56} 1123--1130 

\bibitem{Dick03}
Dickey L A 2003 
{\em Soliton Equations and Hamiltonian Systems} 
(Singapore: World Scientific) 

\bibitem{Gutt+Rawn99}
Gutt S and Rawnsley J 1999 
Equivalence of star products on a symplectic manifold; an
introduction to Deligne's \v{C}ech cohomology classes 
{\em J. Geom. Phys.} {\bf 29} 347--392 

\bibitem{Namb73}
Nambu Y 1973 
Generalized Hamiltonian dynamics 
{\em Phys. Rev. D} {\bf 7} 2405--2412 

\bibitem{Takh94}
Takhtajan L 1994 
On foundations of the generalized Nambu mechanics 
{\em Commun. Math. Phys.} {\bf 160} 295--315 

\bibitem{DFST97}
Dito G, Flato M, Sternheimer D and Takhtajan L 1997 
Deformation quantization and Nambu mechanics 
{\em Commun. Math. Phys.} {\bf 183} 1--22 

\bibitem{ALMY01}
Awata H, Li M, Minic D and Yoneya T 2001 
On the quantization of Nambu brackets 
{\em JHEP} {\bf 02} 013 

\bibitem{Curt+Zach03}
Curtright T and Zachos C 2003 
Classical and quantum Nambu mechanics 
{\em Phys. Rev. D} {\bf 68} 085001

\bibitem{Hiet97}
Hietarinta J 1997
Nambu tensors and commuting vector fields 
{\em J. Phys. A: Math. Gen.} {\bf 30} L27--L33 

\bibitem{Gess84}
Gessel I M 1984 
Multipartite P-partitions and inner products of skew Schur functions 
{\em Contemp. Math.} {\bf 34} 289--301

\bibitem{Reut93}
Reutenauer C 1993 
{\em Free Lie Algebras} 
(Oxford: Clarendon Press) 

\bibitem{Haze00}
Hazewinkel M 2000 
Quasi-symmetric functions 
{\em Formal Power Series and Algebraic Combinatorics} 
ed D Krob, A A Mikhalev and A V Mikhalev 
(Berlin: Springer) pp 30--44 

\bibitem{DNS89}
Dorfmeister J, Neher E and Szmigielski J 1989 
Automorphisms of Banach manifolds associated with the KP-equation 
{\em Quart. J. Math. Oxford} {\bf 40} 161--195 

\bibitem{Macd95}
Macdonald I G 1995 
{\em Symmetric functions and Hall polynomials} 
(Oxford: Oxford University Press) 

\bibitem{Olve+Sand00}
Olver P J and Sanders J A 2000 
Transvectants, modular forms, and the Heisenberg algebra 
{\em Adv. Math.} {\bf 25} 252--283 

\bibitem{Atho07}
Athorne C 2007
Applications of transvectants 
{\em Preprint} (Univ. of Glasgow) 

\end{thebibliography}
\end{document}